\documentclass[lettersize,journal]{IEEEtran}

\title{GlyphWeaver: Unlocking Glyph Design Creativity \\with Uniform GDSL and AI}

\author{%
  Can Liu,
  Shiwei Chen,
  Zhibang Jiang,
  Yong Wang
    \thanks{C. Liu is with the School of Information and the Institute for the Community for the Chinese Nation at the Central University of Finance and Economics, and with Nanyang Technological University.
  This work was conducted primarily while he was a research fellow supervised by Yong Wang at Nanyang Technological University.
  Email: can.liu@cufe.edu.cn.}
  \thanks{S. Chen, and Y. Wang are with Nanyang Technological University. Email: can.liu@ntu.edu.sg, shiwei004@e.ntu.edu.sg, yong-wang@ntu.edu.sg.}
  \thanks{Y. Wang is the corresponding author.}
  \thanks{Z. Jiang is with Providence Health and Services. Email: zhibang.jiang@newschool.edu.}
}

\usepackage{multirow}
\usepackage{tabu}                      
\usepackage{booktabs}                  
\usepackage{lipsum}                    
\usepackage{mwe}                       
\usepackage{hyperref}
\usepackage{xurl}

\usepackage{mathptmx}                  

\usepackage{CJK}            
\usepackage{listings}

\usepackage{enumitem}
\usepackage{needspace}
\usepackage{algorithm}
\usepackage{algorithmic}

\usepackage{amsmath}

\usepackage{soul}
\setstcolor{red}

\newcommand{\icon}[1]{\raisebox{-0.3\height}{\includegraphics[height=1.2em]{image/#1.png}}}

\setstcolor{red}
\ifdefined\showrevisions
  \newcommand{\revision}[1]{\leavevmode{\textcolor[RGB]{0, 102, 204}{#1}}}
  \newcommand{\remark}[1]{\textcolor[RGB]{150, 200, 0}{#1}}
  \newcommand{\removed}[1]{\leavevmode{\color{red}{\st{#1}}}}
\else
  \newcommand{\revision}[1]{#1}
  \newcommand{\remark}[1]{}
  \newcommand{\removed}[1]{}
\fi

\newcommand{\funcname}[1]{\texttt{#1}}

\usepackage{tikz}

\definecolor{basicContainerColor}{RGB}{3, 102, 214}
\newcommand*\basicContainer[1]{\tikz[baseline=(char.base)]{
            \node[shape=rectangle,fill=white, draw=basicContainerColor, text=basicContainerColor, inner sep= 2pt, minimum size=11pt,rounded corners=3pt] (char) {\textbf{#1}}}}

\definecolor{repeatContainerColor}{RGB}{34, 134, 58}
\newcommand*\repeaterContainer[1]{\tikz[baseline=(char.base)]{
            \node[shape=rectangle,fill=white, draw=repeatContainerColor, text=repeatContainerColor, inner sep= 2pt, minimum size=11pt,rounded corners=3pt] (char) {\textbf{#1}}}}

\definecolor{compositeContainerColor}{RGB}{111, 66, 193}
\newcommand*\compositorContainer[1]{\tikz[baseline=(char.base)]{
            \node[shape=rectangle,fill=white, draw=compositeContainerColor, text=compositeContainerColor, inner sep= 2pt, minimum size=11pt,rounded corners=3pt] (char) {\textbf{#1}}}}

\setlist[itemize]{leftmargin=8pt, itemsep=0pt, topsep=2pt}

\newcommand{\toolname}{\textit{GlyphWeaver}}

\usepackage{caption}

\begin{document}

\maketitle

\begin{abstract}
Glyph-based visualizations depict data using collections of glyphs whose visual channels represent data attributes. Authoring such visualizations can require substantial programming effort while making it difficult to retain user control during iterative design. We present \toolname{}, an interactive prototype for authoring glyph-based visualizations while retaining user control over their structures and data mappings. \toolname{} comprises three key components: a glyph domain-specific language (GDSL), a GDSL operation management mechanism, and a multimodal interaction interface. The GDSL is a hierarchical container model that uses \textit{basic}, \textit{repeater}, and \textit{compositor} containers to represent the structures of glyph-based visualizations. The operation management mechanism restricts modifications of the GDSL to five atomic operations, enabling inspectable editing of GDSL instances.
The multimodal interaction interface supports direct manipulation, natural-language commands, and parameter adjustments. A multimodal large language model translates the natural-language commands into GDSL operations. We evaluate \toolname{} through two case studies, a quantitative translation benchmark, and a user study with 13 participants.
The results demonstrate the effectiveness and good usability of \toolname{} in authoring glyph-based visualizations through the benchmark instructions.

\end{abstract}


\begin{CJK}{UTF8}{gbsn}     




\begin{figure*}
    \centering
    \includegraphics[width=\textwidth]{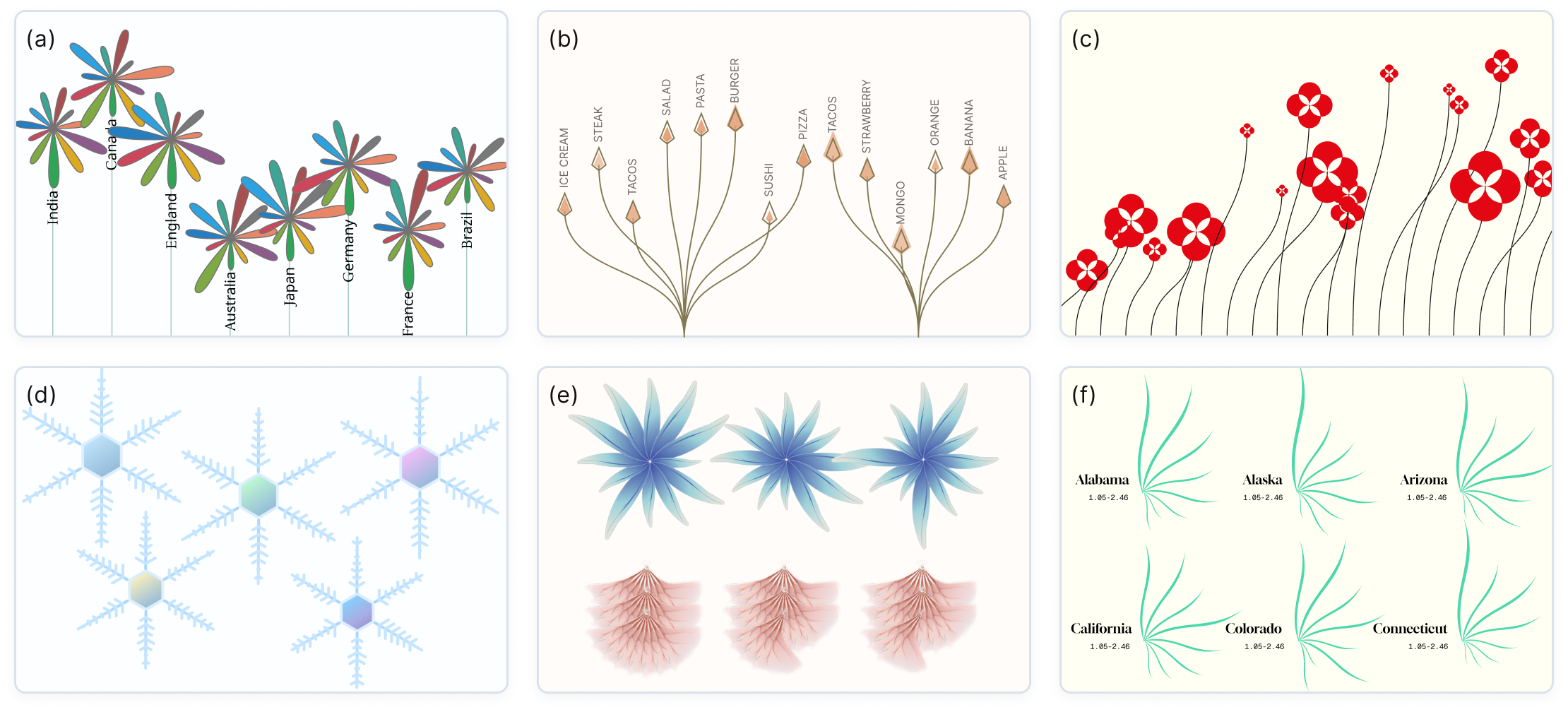}
    \caption{Glyph-based visualization examples recreated with \toolname{}. All the original glyph-based visualizations are award-winning designs from `\textit{Information is beautiful}'\cite{InformationIsBeautifulAwards2025} (a~\cite{stefaner2011oecd}, b~\cite{McCandless2023Protein}, c~\cite{PoppyField}, d~\cite{TwitterInteractive2025}, e~\cite{behanceShapeDreamsGoogle2020}, f~\cite{15minuteInstatePhone}).
    }
    \label{fig:cases}
\end{figure*}

\section{Introduction}








Glyph-based visualization depicts a dataset using a collection of visual objects, referred to as glyphs, whose visual attributes encode one or more attributes of data records~\cite{borgo2013glyph,ward2008multivariate}. Examples appear in award-winning designs from \textit{Information is Beautiful}~\cite{InformationIsBeautifulAwards2025} (\autoref{fig:cases}), which were originally crafted manually using graphics software and later reproduced using our approach. For instance, the OECD example in \autoref{fig:cases}a maps eleven quality-of-life attributes to eleven petals arranged radially around a center~\cite{stefaner2011oecd}. Its structure combines repeated petals with other visual elements in a nested composition. Such structures fall outside the bounded chart design space of high-level grammars such as Vega-Lite~\cite{satyanarayan2017vega}, which targets concise specifications of common statistical graphics.

Despite their value, glyph-based visualizations can be challenging and time-consuming to implement.
The challenges arise from two primary factors: \textbf{the structural complexity and diversity of their glyphs} (\textbf{C1}) and \textbf{the need for human control during the design process} (\textbf{C2}).
The examples in \autoref{fig:cases} use different structures and data mappings; their construction may involve parameterized repetition, nested arrangements, and compositions of heterogeneous elements. Designers often construct such glyphs manually in graphics software such as Adobe Illustrator, Figma, and Sketch.
During this process, designers must specify data-to-visual mappings, organize repeated elements, and refine visual forms and styles.

\begin{figure*}
    \centering
    \includegraphics[width=\textwidth]{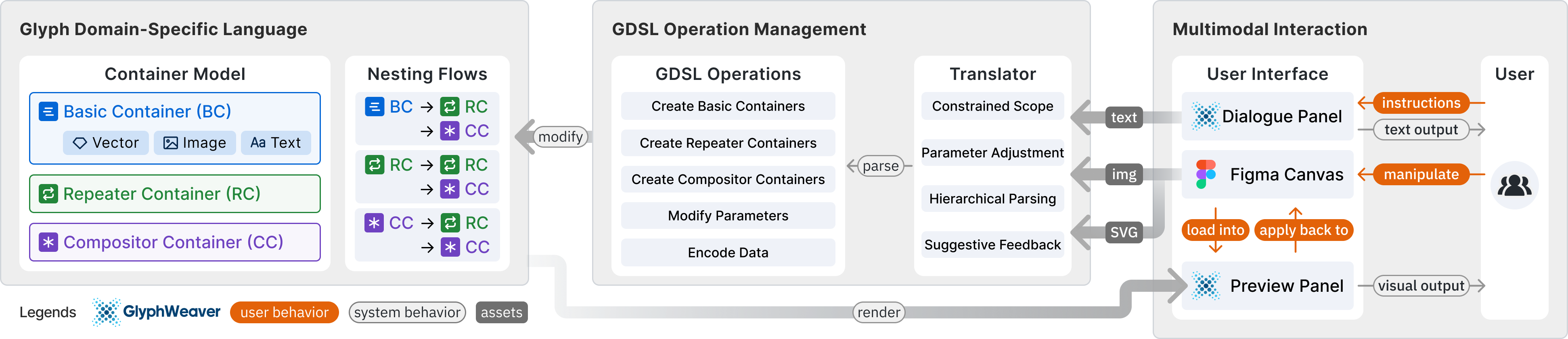}
    \caption{\toolname{} consists of three components:
    The glyph domain-specific language is the theoretical foundation; the GDSL operation management mechanism serves as an intermediary layer between GDSL and the multimodal interaction interface; the multimodal interaction interface allows users to interact with the user interface using natural language and direct manipulation.
    }
    \label{fig:pipeline}
\end{figure*}

Prior research~\cite{metaglyph2023lu,glyphcreator2022lu} has attempted to lower the authoring barrier through automated approaches. MetaGlyph~\cite{metaglyph2023lu} maps tabular data to elements decomposed from a metaphoric image, but does not support nested or hierarchical structures. GlyphCreator~\cite{glyphcreator2022lu} focuses on circular glyph designs, limiting the range ofstructures it can represent. Thus, these approaches reduce manual effort within their respective design spaces but do not cover the heterogeneous and nested structures considered in this work. Low-level programming libraries such as D3.js~\cite{bostock2011d3} provide the flexibility to implement such structures, but require designers to translate visual concepts into code.
Therefore, a gap still exists between flexible low-level implementation and accessible authoring support for glyph-based visualizations.
To address this gap, we propose \toolname{}, an interactive prototype for authoring glyph-based visualizations with reduced manual implementation. Given user-created graphical elements and instructions expressed through natural language or direct manipulation, \toolname{} constructs or updates an editable \revision{Glyph Domain-Specific Language (GDSL)} specification and renders the corresponding glyph-based visualization.
\toolname{} integrates three core components (Fig.~\ref{fig:pipeline}): (1) a \revision{\textbf{GDSL}} with composable containers, (2) a \textbf{GDSL operation management mechanism} based on atomic operations and multimodal large language models (MLLMs), and (3) a \textbf{multimodal interaction interface} supporting both natural language and direct manipulation.
At the core of \toolname{} is the GDSL, 
a structured representation \textbf{(C1)} that uses three container types: basic, repeater, and compositor.
A basic container holds a visual mark (e.g., a bar in a bar chart), a repeater container holds multiple repeated instances of contents based on user-defined parameters, and a compositor container arranges multiple different parts.
%
%
For example, the OECD glyph in \autoref{fig:cases}a can be represented by a basic container for one petal, a repeater container that creates and radially arranges eleven petals, and a compositor container that combines the repeated petals with the center. We derived these container types by analyzing 50 glyph-based visualizations from the \revision{\textit{Information is Beautiful}~\cite{InformationIsBeautifulAwards2025} winning list} and the D3 Gallery~\cite{d3_gallery}; all analyzed structures could be represented by combinations of the three container types.
%
%
%
Then, the GDSL operation management mechanism defines five atomic GDSL operations: creating basic containers, creating repeater containers, creating compositor containers, modifying parameters, and encoding data.
By nesting these atomic operations, different structures can be constructed.
The GDSL operation management mechanism leverages MLLMs to interpret user intentions and translate them into corresponding GDSL operations \textbf{(C2)}.
The multimodal interaction interface accepts natural-language input through the dialogue panel (\autoref{fig:interface}(c)) and direct manipulation on the Figma canvas (\autoref{fig:interface}(a)), providing two ways for designers to specify and revise operations \textbf{(C2)}.
%
%

These three components work together in a coupled workflow: users interact with the prototype through the multimodal interaction interface, expressing their design intentions via natural language or direct manipulation; the operation management mechanism interprets these inputs and translates them into predefined GDSL operations; these operations are then applied to the GDSL to construct or modify the resulting structures.
This workflow keeps MLLM output within the GDSL operation space and exposes inferred parameters for user adjustment.
%
Compared with low-level implementation, \toolname{} is designed to reduce the manual effort required to create glyph-based visualizations while retaining human control.
%
%
We conducted two case studies, a translation benchmark, and a user study to evaluate \toolname{}'s support for interactively authoring glyph-based visualizations.
Our contributions are as follows:
\begin{itemize}
    \item We introduce a hierarchical GDSL for representing the structures of glyph-based visualizations through basic, repeater, and compositor containers, together with five atomic operations for modifying GDSL instances.
    \item We propose \toolname{}, an interactive prototype that incorporates GDSL with a GDSL operation management mechanism to translate natural-language and direct-manipulation input into inspectable operations. We further extensively evaluate it through case studies, a translation benchmark, and a user study with 13 participants.
    
\end{itemize}
We have released glyph-based visualizations created with \toolname{} here: \url{https://glyphweaver.github.io/}.

\section{Related Work}

Our work is related to glyph-based visualization, domain-specific language, and AI-powered visualization generation.

\subsection{Glyph-Based Visualization}

Glyph-based visualizations are compact visual representations that map multiple data variables onto visual channels such as color, shape, orientation, and size~\cite{ward2008multivariate}.
They have long been employed in multivariate visualization~\cite{shaw1999using, post1995iconic}, vector field visualization~\cite{glyphvector1996wittenbrink}, and domain-specific applications such as biological systems~\cite{maguire2012taxonomy} and human-automation collaboration~\cite{nylin2024glyph}.
Foundational surveys synthesize the evolution of glyph-based techniques, design principles, and evaluation methodologies~\cite{borgo2013glyph,fuchs2016systematic}.
Design principles and conceptual models have been proposed to guide effective glyph creation, including perceptual uniformity, redundant encoding, importance ordering, and spatial arrangement~\cite{ward2008multivariate,borgo2013glyph}.
Prior glyph literature further distinguishes one-to-one mappings, in which individual data dimensions are assigned to distinct visual channels, from many-to-one mappings, in which multiple dimensions jointly affect the same visual channel~\cite{ward2008multivariate,borgo2013glyph}.
Prior studies have examined how glyph form and semantic relevance affect interpretation. Flury and Riedwyl~\cite{flury1981asymmetrical} introduced asymmetric facial glyphs for multivariate data. Siirtola~\cite{siirtola2005data} found that data-related glyphs were preferred and produced higher accuracy than unrelated glyphs in the tasks examined.
Li et al.~\cite{li2015metaphoric} further found that metaphoric associations could facilitate interpretation when they were compatible with the represented concept, but could interfere with interpretation when they were incompatible.
%
Together, these findings show that glyph form and semantic relevance can influence interpretation and task performance, although their effects depend on the particular design and task.
Despite their capacity to encode multiple variables in compact forms, glyph-based visualizations with heterogeneous structures remain challenging to design.
Traditional approaches often require manual manipulation or low-level programming~\cite{ribarsky1994glyphmaker}, which imposes a steep technical barrier compared to charting tools like Vega-lite~\cite{satyanarayan2017vega} or ECharts~\cite{li2018echarts}.
Recent approaches have attempted to automate glyph creation within more specific design spaces. GlyphCreator generates circular glyphs from examples~\cite{glyphcreator2022lu}, whereas MetaGlyph generates metaphoric glyphs by decomposing a user-selected metaphoric image and mapping tabular data to its visual elements~\cite{metaglyph2023lu}. These method-specific assumptions limit their support for heterogeneous and nested structures.

\subsection{Domain-Specific Language in Visualization}


Domain-Specific Languages (DSLs)~\cite{mcnutt2022no} provide structured abstractions for visualization design.
Existing systems like Lyra~\cite{satyanarayan2014lyra} and Data Illustrator~\cite{liu2018data} combine direct manipulation with programmatic control, demonstrating how DSLs integrate usability with semantic precision.
Both general-purpose and specialized DSLs have been proposed. General-purpose DSLs such as Vega-Lite~\cite{satyanarayan2017vega} provide concise declarative specifications for a bounded chart design space, whereas specialized DSLs target structures such as trees~\cite{li2020gotree}, networks~\cite{scott2023netpanorama, han2024nuwa}, or multidimensional data~\cite{stolte2002polaris}.
Visualization DSLs span different levels of abstraction: low-level representations expose flexible construction primitives, whereas higher-level declarative grammars provide more concise specifications for a bounded design space~\cite{mcnutt2022no,bostock2011d3,satyanarayan2017vega}.
For instance, D3.js~\cite{bostock2011d3} exposes low-level primitives for custom implementations;
Vega-Lite~\cite{satyanarayan2017vega} sacrifices some flexibility for ease of use via high-level declarative specifications.
Recent research on visualization DSLs has tried to address this tension.
For instance, Bluefish~\cite{pollock2024bluefish} promotes cognitive alignment with declarative abstractions.
Mascot~\cite{liu2024mascot} categorizes operations into generative and modificative primitives
to preserve interactivity while retaining access to low-level operations.
DataGarden~\cite{offenwanger2024datagarden} further attempts to bridge informal sketches with structured templates.
DataSway~\cite{xie2025datasway} generates and coordinates animation clips for SVG-based metaphoric visualizations.

Inspired by the above research, we propose GDSL, which enables the creation of glyph-based visualizations through composition.
GDSL is paired with a set of operations and uses an MLLM to translate user inputs into GDSL operations, embodying the principle of agency plus automation~\cite{heer2019agency} and augmenting human control with AI assistance when creating glyph-based visualizations.

\subsection{AI-powered Visualization Generation}
AI-powered visualization generation spans multiple output representations and levels of user control~\cite{wang2021survey,wang2021ai4vis}, ranging from chart recommendation and specification generation to the direct generation of visual artifacts and code.

Natural-language interfaces such as Articulate, DataTone, NL4DV, FlowSense, and Iris support conversational data analysis or translate user queries into visualizations~\cite{sun2010articulate,gao2015datatone,narechania2020nl4dv,yu2020flowsense,fast2018Iris}. Learning-based approaches similarly generate specifications or recommend chart designs~\cite{dibia2019data2vis,liu2021advisor,luo2021nl2vis,luo2018deepeye,hu2019vizml}. Complementing these learned approaches, Draco formalizes visualization design knowledge as constraints~\cite{moritz2019draco}. KG4Vis, AdaVis, and LLM4Vis further improve recommendation transparency through structured knowledge or natural-language explanations~\cite{li2021kg4vis,zhang2023adavis,llm4vis}. These methods primarily produce chart-level specifications or recommendations rather than editable representations of such structures.

Other systems directly generate infographics, semantically enriched charts, visualization code, or adapted SVG examples~\cite{cui2020text2viz,xiao2023let,tian2024chartgpt,chen2021plotcoder,xie2025datawink}. Although these approaches broaden the range of generated outputs, artifact or code generation alone does not expose the domain-specific operations needed to construct and revise nested glyph components.

\toolname{} therefore uses an MLLM as a translator rather than as an unconstrained end-to-end generator. Its output is restricted to atomic operations over an explicit GDSL, while users can combine natural-language input with direct manipulation and inspect the resulting intermediate states. This design connects AI assistance to a hierarchical representation specialized for the interactive and controllable authoring of glyph-based visualizations.

\section{Overview}


\toolname{} is an interactive glyph-authoring prototype with three components (Fig.~\ref{fig:pipeline}): the glyph
domain-specific language (GDSL) represents heterogeneous and nested structures in glyph-based visualizations and their data mappings to address C1; the operation management mechanism applies five supported GDSL operations to GDSL instances; and the multimodal interaction interface captures user input through natural language and direct manipulation while keeping inferred parameters available for user inspection and adjustment, addressing C2.
This overview provides a roadmap of the architecture; the following three sections describe the GDSL, operation management mechanism, and multimodal interaction interface in detail.

\begin{itemize}[noitemsep,topsep=2pt,parsep=0pt,partopsep=0pt]

\item \textbf{Glyph Domain-Specific Language.}
The GDSL (\autoref{sec:dsl}) is a container-based and structure-driven representation for these structures.
At its core, GDSL treats every visual element as part of a container hierarchy, where containers can hold primitives, repeated elements, or composite assemblies.
This approach allows users to build complex structures by nesting and composing containers.

\item \textbf{GDSL Operation Management Mechanism.}
Directly authoring a GDSL instance requires knowledge of its syntax and container hierarchy.
To address this, we introduce an operation management mechanism (\autoref{sec:state_manage}) that restricts modifications of GDSL instances to five atomic GDSL operations.
While each operation is simple and well-defined, they can be composed to construct complex structures.
Such a mechanism translates user intentions into these operations, reducing the learning curve.


\item \textbf{Multimodal Interaction Interface.}
\toolname{} provides a multimodal interaction interface (\autoref{sec:interaction}) that supports
direct manipulation via the Figma interface, natural language (NL) input through a dialogue panel, and parameter tuning through interactive controls.
Leveraging an MLLM, the framework translates these inputs into GDSL operations.

\end{itemize}

\section{Glyph Domain-Specific Language}
\label{sec:dsl}
\begin{figure*}[ht]
    \centering
    \includegraphics[width=\textwidth]{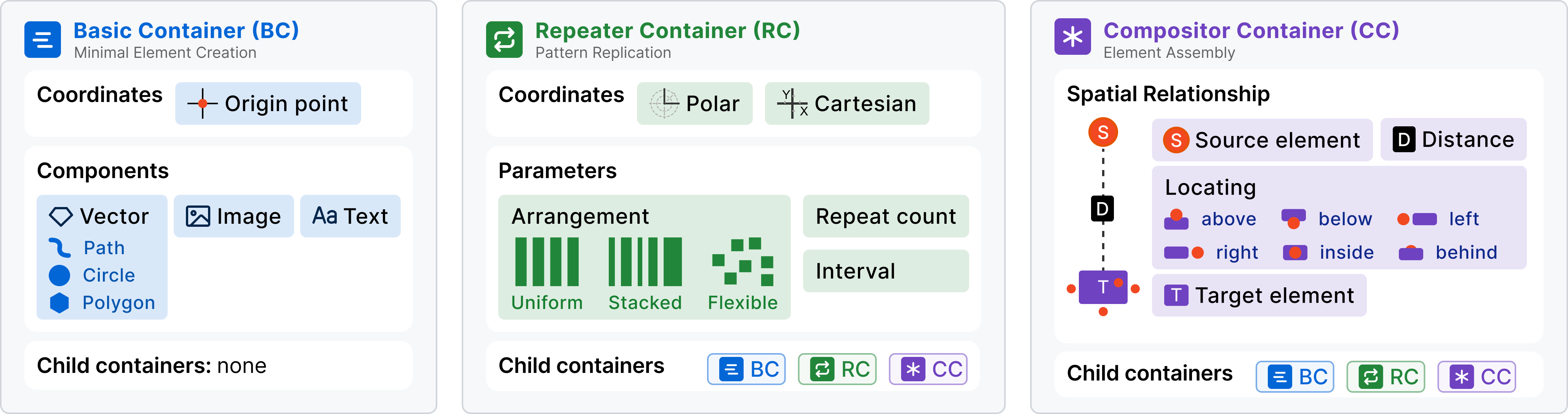}
    \caption{GDSL has three containers:
    Basic containers encapsulate individual visual elements with an independent coordinate.
    Repeater containers replicate a child container using Cartesian or polar coordinates, supporting three arrangements (uniform, stacked, and flexible).
    Compositor containers coordinate spatial relationships between child containers.
    }
    \label{fig:dsl_container}
\end{figure*}

This section introduces the GDSL, a hierarchical container-based model for representing the structures of glyph-based visualizations, as illustrated in \autoref{fig:dsl_container}.
It is comprised of the containers (\autoref{sec:basic_concept}) and
the container relationships (\autoref{sec:hierarchical_structure}).
We also use a GDSL instance to explain its usage (\autoref{sec:gdsl_case}).


\subsection{Containers}
\label{sec:basic_concept}

We analyzed 39 award-winning visualizations from \textit{Information is Beautiful} and 11 D3 Gallery examples, totaling 50 cases.
To derive the structural abstractions used in GDSL, we manually decomposed each glyph-based visualization into its construction process and recorded the recurring structural operations involved in assembling visual elements.
Through iterative comparison across cases, we observed that most construction processes could be consistently interpreted as combinations of three recurring patterns:
(1) constructing individual visual primitives,
(2) duplicating elements with parameterized variation, and
(3) spatially composing heterogeneous structures.
These patterns eventually informed the design of the three container abstractions in GDSL:
basic containers, repeater containers, and compositor containers.

Not every glyph uses all three containers, but their combinations represented all 50 analyzed structures.

All analyzed visualizations employed basic and repeater containers, while 64\% additionally involved compositor structures.
Across the 233 recorded container operations, we identified 89 basic operations, 105 repeater operations, and 39 compositor operations.
This decomposition is also consistent with common visualization construction paradigms such as D3.js, where visual primitives are first instantiated and encoded through data binding, repeated through enter mechanisms, and further organized through hierarchical grouping and composition.
While the analyzed examples are not exhaustive, they cover a broad range of structures found in infographic and visualization practice, including radial layouts, repeated structures, and nested assemblies.

\begin{itemize}[noitemsep,topsep=2pt,parsep=0pt,partopsep=0pt]
\item \basicContainer{Basic containers (BC)} encapsulate basic elements (polygons, lines, images, and text).
A \basicContainer{BC} maintains an independent coordinate system.
It enables precise control over visual attributes such as position, scale, rotation, and color, making it a fundamental building block in glyph design.
Basic containers typically encode data in the form of a dictionary, where each data value is directly mapped to the visual attributes of an element.
For example, a number value can be mapped to the width and height of a rectangle.

\item \repeaterContainer{Repeater containers (RC)} replicate a unit multiple times using either the Cartesian or polar coordinate systems.
Cartesian coordinates are for grid-based and linear arrangements, and polar coordinates are for circular arrangements.  
A \repeaterContainer{RC} contains a child container and defines both the repetition count and arrangement pattern, replicating the child container into multiple instances according to this specified arrangement.
An \repeaterContainer{RC} encodes a list of data values, where each item is mapped to an instance of its child container.
For example, if a petal is rotated and duplicated 12 times to form a flower, and different petals are expected to have varying sizes, each entry in the data array would contain a unique scale value.

\item \compositorContainer{Compositor containers (CC)} integrate multiple different child containers.
\compositorContainer{CCs} also support both Cartesian and polar coordinate systems.
Unlike \repeaterContainer{RCs} that manage duplications of a child container, \compositorContainer{CCs} coordinate the positioning of different child containers.
A \compositorContainer{CC} typically use dictionary-structured data, where different keys correspond to different child containers.
\end{itemize}

\textbf{Spatial layout of containers.}
The positions of containers are influenced by their inner coordinate system, the transformation relative to the parent, and the spatial relationship between sibling containers.

\begin{figure}[htp]
    \centering
    \includegraphics[width=\columnwidth]{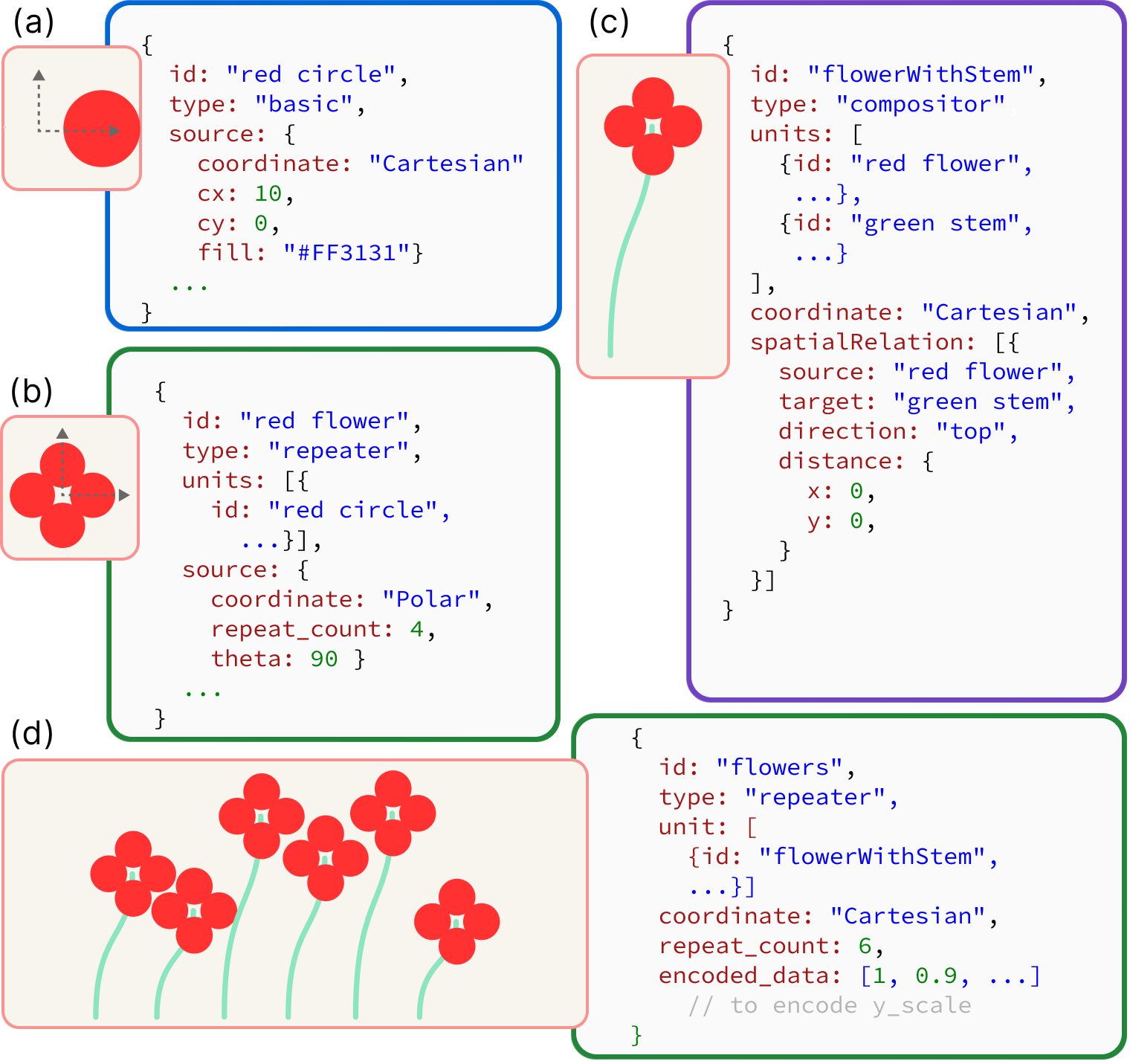}
    \caption{
    A GDSL instance with basic container (a), repeater containers (b and d), and compositor container (c).}
    \label{fig:dsl_instance}
\end{figure}

\begin{itemize}[noitemsep,topsep=2pt,parsep=0pt,partopsep=0pt]
    \item \textbf{Inner coordinate system.}
    Each container (\basicContainer{BC}, \repeaterContainer{RC}, and \compositorContainer{CC}) possesses an independent coordinate system for positioning its internal elements with an origin point.
    These coordinate systems can be either Cartesian \icon{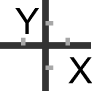} or polar \icon{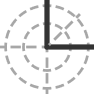}.
    
    \item \textbf{Parent-relative transformation.}
    This property defines the transformation of a container relative to its parent container, including three types: translation, rotation, and scaling.
    The transformation is assigned by the parent container or independently mapped using the container's own encoded data.
    In \repeaterContainer{RCs} (\autoref{fig:dsl_container}), for instance, the parent can assign different transformers to its child repeat instances, as shown in \autoref{fig:dsl_instance} (b), where the red flower assigns varying rotation angles to four circles.
    There are three repetition arrangements: uniform, stacked, and flexible, as shown in \autoref{fig:dsl_container}:
    Repetitive instances in uniform arrangement \icon{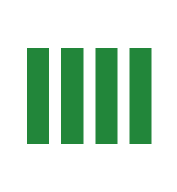} are placed at uniform intervals, ensuring consistent spacing.
    In a Cartesian system, intervals are defined by $x$ and $y$ values; in a polar system, by radius and angle.
    Repetitive instances in stacked repetition \icon{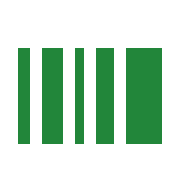} are recursively positioned, where each element's position is determined by previous instances.
    Repetitive instances in flexible repetition \icon{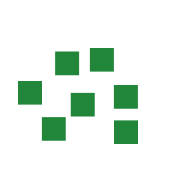} are arranged with variable spacing or transformations, allowing for more organic and dynamic layouts.
    A container in a flexible arrangement can determine its own transformer based on data, as illustrated in \autoref{fig:dsl_instance} (d), where green stems scale in the y-direction according to their individual data, resulting in flowers of diverse heights.
    
    \item \textbf{Sibling container relationship.}
    Within \compositorContainer{CCs}, spatial relationships may exist between sibling containers.
    Spatial relationships in \toolname{} are defined as distances between anchor points on source \icon{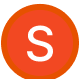} and target \icon{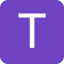} (\autoref{fig:dsl_container} \compositorContainer{CC}).
    Each spatial relationship can define anchor points as reference points, which may correspond to geometric features of the container (such as center, edge, or corner points).
    The distance \icon{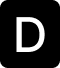} (as a 2D vector) between the anchor points of the source and target containers is also given.
    As shown in \autoref{fig:dsl_instance} (c), a spatial relationship exists between the red flower and the green stem, with the red flower anchored to the top point of the green stem with a distance of $0$.
    When calculating positions, these spatial relationships are translated into constraints (equations based on anchor points and distances), which are solved to determine the final (x, y) coordinates of all containers.
\end{itemize}

\subsection{Hierarchical Container Relationships}
\label{sec:hierarchical_structure}

GDSL employs a hierarchical container system that enables nested structure while maintaining manageable complexity.

\textbf{Cognitive alignment.}
A hierarchical structure inherently aligns with users' cognitive processes.
Users typically begin with basic visual marks and progressively develop complexity by applying repetition and composition.
Complex visual structures can be derived through the orchestration of three container types.
This hierarchical architectural framework not only facilitates extensibility but also promotes the development of reusable components that can be leveraged in subsequent design iterations.

\textbf{Nested construction.}
The construction process in GDSL leverages this hierarchical approach through a nesting mechanism.
\basicContainer{BCs} serve as the foundation, which can then be duplicated to form \repeaterContainer{RCs}, while multiple different containers can be combined into a compositor container.
This process can continue recursively: a compositor container itself can  be a child of a repeater container.

\textbf{Complexity management.} 
To manage the potential complexity that arises from such hierarchical nesting, GDSL implements several control mechanisms.
GDSL encapsulates parameters and transformations at each level of the container hierarchy.
Each container operates within its independent coordinate system, preventing spatial transformations from interfering across different levels.
Additionally, GDSL maintains independent data management at each hierarchical level, preserving a clear data flow throughout the structure.
Each level also maintains a separate management system that governs parameter mapping rules specific to its scope.
This hierarchical and modular approach allows GDSL to maintain clarity and control while enabling users to create complex visualizations through element composition and repetition.

\subsection{GDSL Instance Example}
\label{sec:gdsl_case}

\begin{table*}[ht]
\centering
\footnotesize
\setlength{\tabcolsep}{3pt}
\caption{The five atomic GDSL operations: three create containers, one modifies parameters, and one maps data to visual attributes.}
\begin{tabular}{|p{1.8cm}|p{2.4cm}|p{5.7cm}|p{5.8cm}|}
\hline
\multicolumn{1}{|c|}{\textbf{Operation Type}} & \multicolumn{1}{c|}{\textbf{Parameters}} & \multicolumn{1}{c|}{\textbf{Example}} & \multicolumn{1}{c|}{\textbf{Explanation}} \\
\hline
\begin{tabular}[c]{@{}p{1.8cm}@{}}\basicContainer{BC} Creation\end{tabular} & 
\begin{tabular}[c]{@{}p{2.2cm}@{}}
id, type, params, coordinate
\end{tabular} & 
\begin{tabular}[c]{@{}p{5.35cm}@{}}
\texttt{CreateBasic("rect1", "rect", \{width, height, fill\}, \{origin, transform\}})
\end{tabular} & 
\begin{tabular}[c]{@{}p{5.45cm}@{}}
Creates a container for a basic graphic element (polygon, line, rect, circle, text) with specified visual attributes.
\end{tabular} \\
\hline
\begin{tabular}[c]{@{}p{1.8cm}@{}}\repeaterContainer{RC} Creation\end{tabular} & 
\begin{tabular}[c]{@{}p{2.2cm}@{}}
id, target\_id, coordinate, repeat\_count, repeat\_type, repeat\_interval
\end{tabular} & 
\begin{tabular}[c]{@{}p{5.35cm}@{}}
\texttt{CreateRepeater("flower", "petal", "polar", 12, "uniform")}
\end{tabular} & 
\begin{tabular}[c]{@{}p{5.45cm}@{}}
Create a container by repeating a child container in polar or Cartesian coordinates.
\end{tabular} \\
\hline
\begin{tabular}[c]{@{}p{1.8cm}@{}}\compositorContainer{CC} Creation\end{tabular} & 
\begin{tabular}[c]{@{}p{2.2cm}@{}}
id, components,\\ spatial\_relations
\end{tabular} & 
\begin{tabular}[c]{@{}p{5.35cm}@{}}
\texttt{CreateCompositor("chart", ["title", "bars"], [\{source, target, type, distance\}])}
\end{tabular} & 
\begin{tabular}[c]{@{}p{5.45cm}@{}}
Combines multiple child containers into a composite structure with spatial relationships.
\end{tabular} \\
\hline
\begin{tabular}[c]{@{}p{1.8cm}@{}}Modify Parameters\end{tabular} & 
\begin{tabular}[c]{@{}p{2.2cm}@{}}
target\_id, params
\end{tabular} & 
\begin{tabular}[c]{@{}p{5.35cm}@{}}
\texttt{ModifyParams("rect1", \{fill: "red", width: 150\})}
\end{tabular} & 
\begin{tabular}[c]{@{}p{5.45cm}@{}}
Modifies static parameters of a container without data mapping, including visual attributes or structural properties.
\end{tabular} \\
\hline
\begin{tabular}[c]{@{}p{1.8cm}@{}}Encode Data\end{tabular} & 
\begin{tabular}[c]{@{}p{2.2cm}@{}}
target\_id, attribute, data\_list | data\_expression
\end{tabular} & 
\begin{tabular}[c]{@{}p{5.35cm}@{}}
\texttt{EncodeData("bars", "height", [10, 45, 30])}
\end{tabular} & 
\begin{tabular}[c]{@{}p{5.45cm}@{}}
Maps data values to parameters or visual attributes, enabling different visual representations based on data values.
\end{tabular} \\
\hline
\end{tabular}
\label{tab:dsl_operations}
\end{table*}

\autoref{fig:dsl_instance} showcases four levels of GDSL instances.
The \basicContainer{BC} \icon{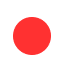} represents the fundamental building unit, as illustrated in \autoref{fig:dsl_instance} (a).
It defines a red circle in a Cartesian coordinate.
Its origin point is on the left of the circle.
The \basicContainer{BC} (with id ``red circle'') from \autoref{fig:dsl_instance} (a) is repeated four times, with each instance rotated by $90$ degrees, resulting in a red flower, the \repeaterContainer{RC} \icon{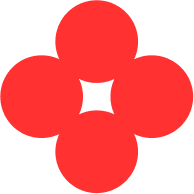} in \autoref{fig:dsl_instance} (b).
The \compositorContainer{CC} in \autoref{fig:dsl_instance} (c) is employed to integrate multiple distinct containers and may define their spatial relationships.
The \compositorContainer{CC} with ID ``flowerWithStem'' integrates two elements, the red flower \icon{four_circle} and the green stem \icon{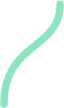}.
The \compositorContainer{CC} establishes their spatial relationship:
the red flower is attached to the top \icon{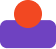} anchor of the green stem with a displacement of $\{x: 0, y: 0\}$.
The \repeaterContainer{RC} in \autoref{fig:dsl_instance} (d) demonstrates a more complex application where the repeating unit itself is a compositor container (with id ``flowerWithStem'').
This unit is repeated six times in a Cartesian coordinate system, with the \texttt{encoded\_data} array providing different scalar values for the y-directional scaling of each ``green stem'' instance, thereby creating a floral arrangement with varying heights.

\section{GDSL Operation Management Mechanism}
\label{sec:state_manage}

%
To enable controllable editing of GDSL instances, 
we propose a GDSL operation management mechanism.
It employs concise and semantically clear operations to ensure structural validity and enhance the editing accuracy and consistency.
This mechanism consists of:
(1) a set of well-defined atomic operations, which serve as the building blocks for GDSL editing; and
(2) an MLLM-powered operation translator, which translates user intentions into valid atomic operations.
%


\subsection{Atomic Operations}  
\label{sec:dsl_operations}

GDSL constructs visual structures through five atomic GDSL operations (\autoref{tab:dsl_operations}), grouped into container creation and modification.
These composable semantic units build a structural skeleton and modify its parameters or bind data.

\begin{figure}[h]
    \centering
    \includegraphics[width=\columnwidth]{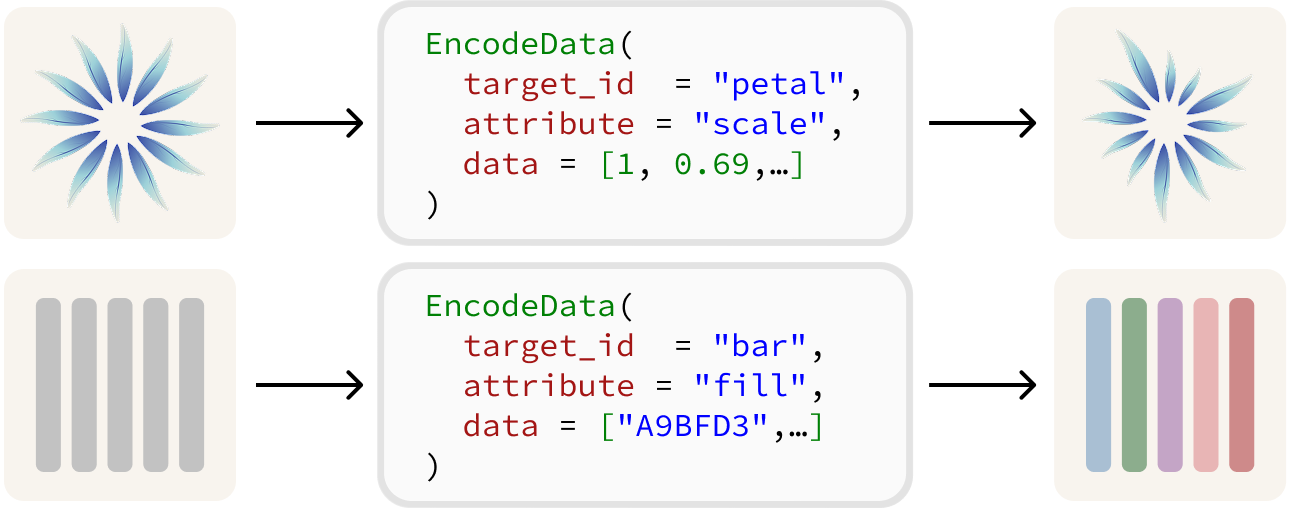}
    \caption{The data encoding operation enables encoding various data items into visual attributes of different instances.}
    \label{fig:encode_data}
\end{figure}

\begin{itemize}[noitemsep,topsep=2pt,parsep=0pt,partopsep=0pt]

\item \textbf{\basicContainer{BC} creation.}
\basicContainer{BCs} encapsulate elements, including polygons, lines, rectangles, circles, and text.
The operation \funcname{CreateBasic} defines the visual attributes for the basic element (e.g., position, size, color), with a coordinate system specification that includes an origin point and transformations (rotation, translation, and scaling).

\item \textbf{\repeaterContainer{RC} creation.}
The \funcname{CreateRepeater} operation replicates existing content to generate a new \repeaterContainer{RC}.
\funcname{CreateRepeater} requires specifying the repeating unit's ID (an existing container), the repetition count, the coordinate type, and the repetition category (uniform, stacked, or flexible).
Figure~\ref{fig:dsl_instance} illustrates a sequence of operations on GDSL instances. The \basicContainer{BC} containing one red circle in panel (a) serves as the repeating unit. Applying \funcname{CreateRepeater} with this target, a repetition count of four, and a uniform polar arrangement produces the four-circle \repeaterContainer{RC} in panel (b). Panel (c) then uses \funcname{CreateCompositor} to attach the resulting flower to a stem, and panel (d) repeats this composite unit six times and binds data to the stems' vertical scales.
%

\item \textbf{\compositorContainer{CC} creation.}
\funcname{CreateCompositor} operation combines multiple existing units into a \compositorContainer{CC}.
Parameters include the information to be combined and potential spatial relationships between components. For each spatial relationship, the source and target IDs, the type of spatial dependency (e.g., top, bottom, left, right), and the spatial distance must be specified. For example, positioning container A above container B with a distance of 5 units would be expressed as:
\texttt{\{source ID = "container A"},  
\texttt{target ID = "container B"},  
\texttt{spatial type = "top"},  
\texttt{distance = }$\{x: 0, y: 5\}\}$.

\item \textbf{Modifying parameters.}
The operation \funcname{ModifyParams} alters the parameters of a target container.
This operation requires providing the target container, visual attribute, and new value.
For example, changing the repetition count of a \repeaterContainer{RC} (with id ``repeat''), the parameters are:
\texttt{\{target\_ID = "repeat"},
\texttt{attribute = "repeat count"},
\texttt{value = 5\}}.

\item \textbf{Encoding data.}
The operation \funcname{EncodeData} enables data attributes to be mapped to visual attributes.
This operation adjusts the parameters of target elements in batches.
Unlike the \funcname{ModifyParams} operation, \funcname{EncodeData} allows different instances to reflect distinct data values.
For example, as shown in \autoref{fig:encode_data}, binding the scale of petals to a data list can transform a flower with uniformly sized petals into one with petals of varying sizes; similarly, binding the fill color of rectangles to a data list can result in rectangles displaying different colors.

\end{itemize}

\subsection{Data Pipeline}

Data can come from user-provided values or a specified simulation function. GDSL organizes the resulting values according to the container hierarchy. Each \texttt{repeat} level introduces a corresponding array dimension, so the data dimensionality must match the target repeated structure. The data-binding component checks this correspondence; when the dimensions do not match, \toolname{} displays a mismatch warning and asks the user to revise the values or simulation function to the expected array shape.
For example, if one petal is replicated into 12 petals, the resulting data structure is an array of size $1 \times 12$. If the flower is then further replicated into three flowers, the data structure becomes an array of size $3 \times 12$.
To support such arrays, we allow users to specify data in two ways:

\begin{itemize}[noitemsep,topsep=2pt,parsep=0pt,partopsep=0pt]
\item \textbf{Math Expressions.}
Users may provide a simulation function that generates data based on the variables \texttt{index} and \texttt{random}, allowing the construction of specific trends or random patterns.
By composing these variables, the MLLM can generate a wide range of mathematical expressions, such as \texttt{(index * 5 + random())} for a sequence with a linear trend and random variation. Synthetic data can also be specified through natural language descriptions, such as \textit{``a dataset with an increasing trend''} or \textit{``random values between 10 and 20''}. The translator then converts these descriptions into mathematical expressions. \toolname{} supports two variables, \texttt{index}, which represents the sequence number of each data point, and \texttt{random}, which generates a random value in $[0,1]$, together with basic arithmetic operations, such as $+$, $-$, $\times$, and $/$.

\item \textbf{Direct Data Provision.}
Beyond expression-based generation, \toolname{} supports binding user-provided or updated lists of values. For example, for text elements in a \repeaterContainer{RC}, a user can bind a list of U.S. state names to the text attribute. The user may request that the translator generate different state names or explicitly provide a list, such as California, Arizona, and Texas.
Users may provide such data either inline or through file upload. For inline input, a user may paste an array into the chat box and specify its target mapping, such as using a one-dimensional array to control petal sizes. \toolname{} parses the input, identifies the corresponding target elements and attributes, such as \texttt{scale}, and then binds the data accordingly. Alternatively, users may upload data files in \texttt{CSV} or \texttt{JSON} format. The corresponding data processing workflow remains consistent with that of chat-based input.
\end{itemize}

\subsection{MLLM-Powered Translator}
\label{sec:mllm_state}

To ensure diverse inputs are translated into GDSL operations in a controllable manner, \toolname{} uses a four-stage translation strategy.
First, \toolname{} constrains the parsing scope.
Within the controlled space, users can interactively adjust parameters to align results with their intent. \toolname{} then supports hierarchical parsing for complex designs, decomposing multi-level structures step by step. Finally, \toolname{} provides suggestive feedback for irrelevant inputs, ensuring even off-topic commands are redirected productively.
In our implementation, natural-language translation is powered by GPT-4o or GPT-5.1, using a structured prompt template to enhance output consistency.
However, we still cannot guarantee that the parsing will always perfectly match the user's intent.
To address this, we provide an undo button as well as ambiguity widgets that allow users to adjust the parsing results (as shown in \autoref{fig:interface}).


%


\begin{itemize}[noitemsep,topsep=2pt,parsep=0pt,partopsep=0pt]

\item \textbf{Constrained Parsing Scope.}
The translator first enforces a restricted output space: the MLLM is explicitly instructed to generate only five supported atomic operations (defined in \autoref{sec:dsl_operations}).
This scope prevents unsupported or overly broad outputs and guarantees predictability.
When user instructions are underspecified but still valid, \toolname{} supplements them with default values. For instance, the input \textit{``rotate and replicate the object six times''} is interpreted as six rotations of $60^{\circ}$ each, forming a circular pattern.
%

\item \textbf{Interactive Parameter Adjustment.}
Even when an instruction maps to one of the five supported atomic operations, its inferred parameters may not align with user intent.
The dialogue panel therefore displays parameters such as angle, scale, and repetition count in ambiguity widgets for direct adjustment (see \autoref{sec:dialog_panel}).
%

\item \textbf{Hierarchical Parsing for Complex Designs.}
For inputs describing multi-level or composite structures, \toolname{} applies hierarchical parsing.
The translator first identifies higher-level structural patterns and then recursively analyzes their components. Each step targets a container structure.

\item \textbf{Suggestive Feedback for Irrelevant Inputs.}
When the input cannot be mapped to one of the five supported GDSL operations, the dialogue panel identifies it as unsupported and prompts the user to reformulate it as a supported glyph-authoring operation.
For example, for the user input \textit{``play a song''} that is not relevant to glyph design or creation, it can return suggestions like \textit{``try `replicate the shape five times' instead.''}

\end{itemize}

For an NL instruction within the supported operation space, the translator returns an atomic operation and its inferred parameters. When a parameter is missing or ambiguous, the dialogue panel exposes the inferred value in an ambiguity widget (\autoref{sec:dialog_panel}). If an instruction cannot be mapped to a valid GDSL operation, the dialogue panel reports that limitation and provides an example of a supported instruction.

\section{Multimodal Interaction Interface}
\label{sec:interaction}

\begin{figure}[h]
    \centering
    \includegraphics[width=\linewidth]{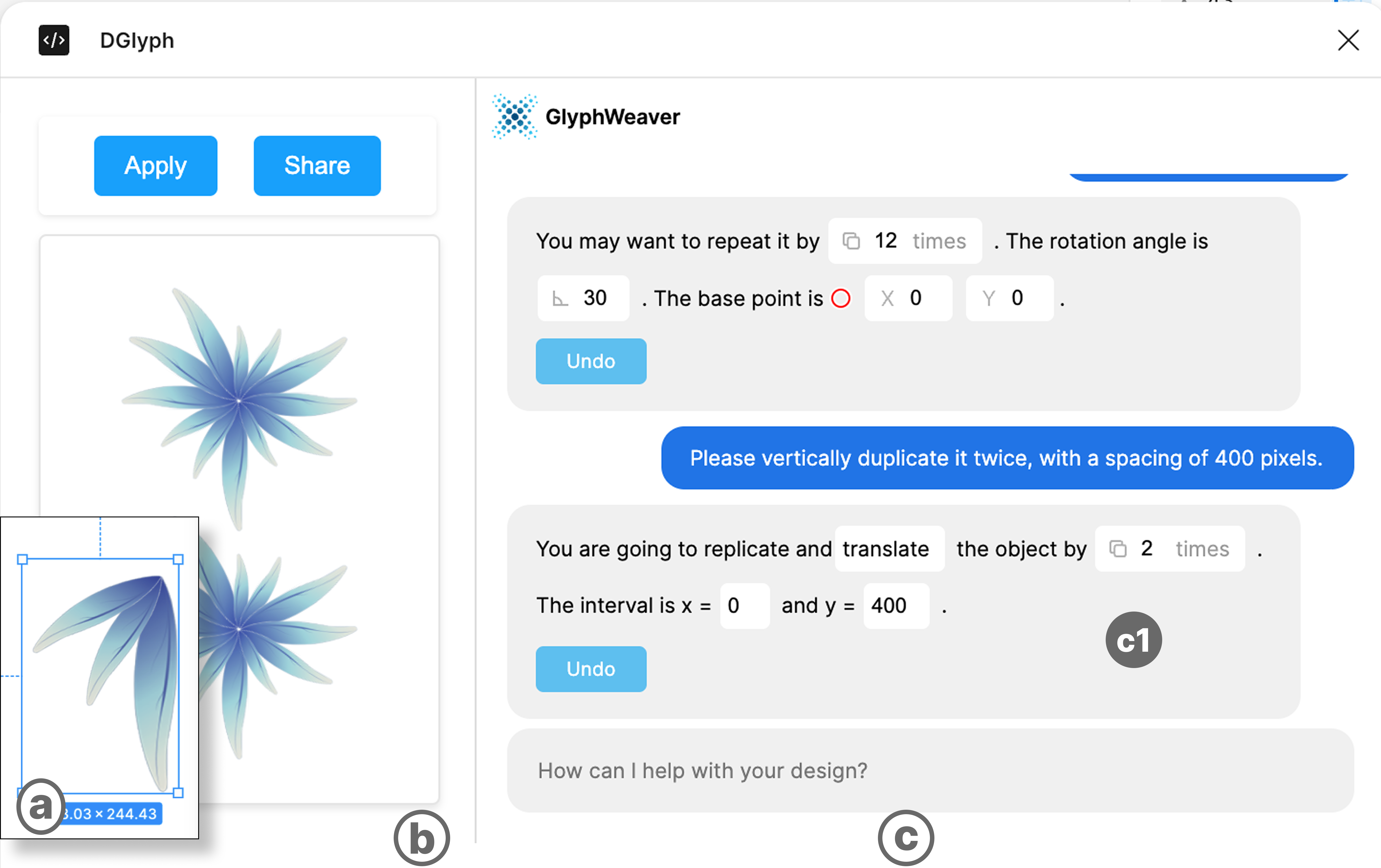}
    \caption{The interface of \toolname{}. (a) the manipulation interface (Figma canvas); (b) the preview panel; (c) the dialogue panel. In the Figma canvas, users can create and edit graphics. In the dialogue panel, users can present textual instructions or modify parameters in the widgets. Users can view current rendered results in the preview panel.}
    \label{fig:interface}
\end{figure}

We implemented \toolname{} as a Figma\footnote{Figma: \url{www.figma.com}} plugin, given Figma's popularity in graphic design.
\toolname{} has three views: the direct manipulation interface (Figma canvas), dialogue panel, and preview panel, as shown in \autoref{fig:interface}.
Users can interact with \toolname{} in a multimodal manner, using direct manipulation or natural language.
Users can directly manipulate graphics in the Figma canvas (\autoref{fig:interface}a).
In the dialogue panel (\autoref{fig:interface}c), users can input natural language (NL) instructions and adjust parameters via ambiguity widgets (inline editable input boxes embedded in the dialogue text \autoref{fig:interface}c1).
Users can view the rendered result and adjust parameters directly in the preview panel (\autoref{fig:interface}b).

\subsection{Direct Manipulation Interface}




We integrate \toolname{} with Figma via the Figma API to leverage Figma's direct manipulation interface.
Users can select graphics directly on the Figma canvas, and \toolname{} monitors these selection changes in real time.
When the selection is updated, \toolname{} prompts the user in the interface: \textit{``You have selected [object]. Would you like to start the parsing process?''} The selected graphics are then converted to SVG and, upon confirmation, parsed into GDSL containers.

%


\begin{itemize}[noitemsep,topsep=2pt,parsep=0pt,partopsep=0pt]
\item \textbf{\basicContainer{BC} Creation}.
For selected elements, \toolname{} creates \basicContainer{BCs} in the GDSL.
Each basic container encapsulates a graphical element, including text, rectangles, circles, paths, or grouped elements treated as a single SVG unit.

\item \textbf{\repeaterContainer{RC} Creation.}
We examine whether any of these \basicContainer{BCs} form repeating patterns. Candidate repeated elements are identified using an MLLM, followed by a rule-based pipeline that infers the geometric transformations relating these elements. This pipeline analyzes the spatial coordinates of repeated elements and organizes them into a hierarchical framework of 2D affine transformations.
The space of affine transformations comprises translation, rotation, scaling, and their combinations (shearing is rarely encountered in visual design tools and is therefore excluded). Accordingly, we focus on translations, rotations around a fixed center, and scalings, either overall ($s$) or along a single direction ($s_x, s_y$). To align with user intent, we adopt a simple-to-complex inference strategy: transformations are fitted in order of increasing complexity, translations first, then single-center rotations, and finally uniform or non-uniform scalings if necessary.
Based on the inferred parameters, we construct \repeaterContainer{RCs} in either Cartesian or polar coordinates, associating scaling parameters with the corresponding data encoding. Elements for which no valid transformation can be inferred remain in their \basicContainer{BCs}.

\item \textbf{\compositorContainer{CC} Creation.}
If the selection contains more than one container, we create a \compositorContainer{CC} to group them.
We position these containers based on their coordinates.
\end{itemize}



\begin{figure}[h]
    \centering
    \includegraphics[width=\columnwidth]{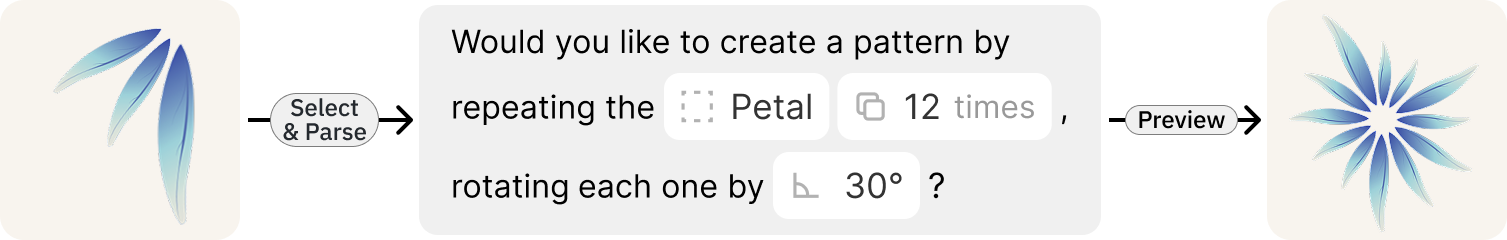}
    \caption{Repeater container creation from Figma canvas.
    Users copy three repetitive petals in the Figma canvas.
    \toolname{} infers that the user intends to create a repeater container with a polar coordinate system.
    The user can further adjust the parameters of the repeater container in the dialogue panel.}
    \label{fig:case_implicit}
\end{figure}


As shown in \autoref{fig:case_implicit}, \toolname{} captures user intent by recognizing qualitative and quantitative parameters from the MLLM-based container analysis.
\toolname{} automatically identifies repeated or rotatable elements and infers minimal transformations.
For repeated petals, \toolname{} automatically identifies a polar arrangement and infers a 30-degree default rotation interval.
Based on the calculated transformations (repetition and rescaling), we represent user intent using operations for the GDSL.
We also provide NL explanations of the parsing results, as displayed in the dialogue panel of \autoref{fig:case_implicit}, allowing users to understand the glyph-authoring actions.


\subsection{Dialogue Panel}
\label{sec:dialog_panel}

Users can input NL commands in the dialogue panel.
Each command is processed by the MLLM translator (e.g., GPT-4o), which takes as input the current GDSL, other existing GDSLs on the canvas, and the user's NL input.
The translator first classifies the instruction into one of four operations (except for creating a basic container) and then extracts the relevant parameters required for execution.



\begin{itemize}[noitemsep,topsep=2pt,parsep=0pt,partopsep=0pt]
\item \textbf{Parameter Modification.}
Users can modify the parameters of a container by setting a single value on a visual attribute.
For example, users may say, \textit{``Change the circle's fill to blue.''}
The translator identifies the target element ID and specified attribute, after which the operation management mechanism applies the update.

\item \textbf{Data Encoding.} Users can provide a list of values, or specify an expression.
For instance, \textit{``Randomize petal sizes between 1 and 1.5.''} prompts the model to generate a data expression.

\item \textbf{\compositorContainer{CC} Creation.}
Users can integrate existing DSL instances from the canvas into the current GDSL, specifying the spatial relationship.
For example, \textit{``Add a triangle above the square''} or \textit{``Place a circle 50 units above the rectangle.''}
The translator determines the referenced element, placement rule (e.g., ``top''), and horizontal or vertical offset before the operation management mechanism applies the composition.

\item \textbf{\repeaterContainer{RC} Creation.}
Users can construct \repeaterContainer{RCs} by giving NL instructions that describe repetition patterns.
For instance, in \autoref{fig:interface}, a user may say, \textit{``Vertically duplicate it twice''}.
\toolname{} determines the target element, repetition count, and the parameters for translation or rotation.
\end{itemize}

\noindent
When an instruction cannot be mapped to an operation, the dialogue panel suggests a supported form, such as \textit{``Specify the element and color, e.g., `Change the circle fill to blue'.''}
By leveraging an MLLM to translate natural language into precise GDSL operations, \toolname{} supports flexible interactions.



\textbf{Ambiguity Widgets.}
Ambiguity widgets~\cite{gao2015datatone} are inline editable boxes (\autoref{fig:interface}c1) developed to expose and resolve ambiguities in NL commands by allowing users to edit parameters.
\toolname{} embeds ambiguity widgets within the dialogue panel's textual feedback, enabling direct parameter adjustment by users.
Users can modify:

\begin{itemize}[noitemsep,topsep=2pt,parsep=0pt,partopsep=0pt]
\item \textbf{Target element selection.}
For each operation, \toolname{} infers a default target container ID from the current GDSL and displays it in the widget.
The widget also provides a drop-down list of all valid IDs, allowing users to select a different target element.


\item \textbf{Numerical values.} For parameters expressed as numeric values (e.g., repetition counts, angles, translation distances, or coordinates), users can edit these values via the widgets.

\item \textbf{Color selection.} Users can refine color attributes by selecting or adjusting colors interactively.
\item \textbf{Data expressions and arrays.} More generally, parameters represented as strings, including generated data expressions and data arrays, are editable as free-form string input.
\end{itemize}
For example, in \autoref{fig:case_implicit}, the widgets displayed alongside the translator response allow users to modify the repeated unit and repetition count.
Adjustments made through ambiguity widgets are immediately reflected in both the GDSL representation and the preview panel, providing instant feedback for refinement.


\subsection{Preview Panel}



\toolname{} provides a preview panel that is synchronized with the current GDSL instance.
The Preview panel acts as a renderer for \toolname{}, where the DSL is instantiated layer by layer according to the rules of repetition, composition, spatial relationships, and data mappings.
The basic elements are represented using SVG primitives such as \texttt{<rect>}, \texttt{<circle>}, and \texttt{<path>} (for polygons and lines), while images are rendered using \texttt{<rect>} with texture filters (\texttt{<filter>}).
Repeating and compositional structures are expanded into corresponding \texttt{<g>} (group) elements, each with appropriate transformation parameters. The instantiation process unfolds the hierarchical GDSL structure in a top-down manner, 
recursively expanding \repeaterContainer{RCs} based on their parameters.
As a result, the rendering time complexity is proportional to the total number of instantiated SVG groups, which is determined by the product of repetition counts across all nested levels in the GDSL. Each repeated structure maps directly to an SVG \texttt{<g>} element, so the overall complexity matches that of standard SVG rendering. While the actual rendering time may be slightly higher due to the overhead of recursively expanding the GDSL structure, the asymptotic complexity remains unchanged.
These groups and basic elements are then rendered in SVG, as shown in \autoref{fig:interface}.

Once the user is satisfied with the rendered result, clicking the Apply button will send the SVG result onto the Figma canvas.
This ensures synchronization among the GDSL instance, the SVG representations in \toolname{} interface, and the graphics in the Figma canvas.
The rendered result synchronized to the Figma canvas is cached on the server via unique identifiers.
In subsequent sessions, users can reuse previous GDSL instances.
For example, if a user previously created a blue flower and later draws a set of lines, the composition algorithm can associate the blue flower with these lines, such as \textit{``put a blue flower on top of each line.''}

\subsection{Interaction Pipeline and Latency}


%
The interaction pipeline exports selected Figma graphics as SVG,
passes the selected SVG representation from the Figma plugin to \toolname{}'s parsing pipeline,
invokes the online MLLM when translation is required, and renders the resulting GDSL instance in the preview panel. \autoref{tab:llm_performance} reports model latencies under the quantitative-evaluation setup in \autoref{sec:quantitative_evaluation}; because response time varies with the model, network conditions, and input modality, we do not claim a general real-time bound.

\section{Evaluation}


We evaluate \toolname{} through two case studies, a translation benchmark, and a user study to assess its performance and usability.
We do not perform a direct quantitative comparison with general-purpose design tools such as Figma, as these systems do not natively support structured glyph repetition, data binding, and natural-language-driven editing. Instead, our evaluation focuses on the usability and interaction effectiveness of the proposed workflow.


\subsection{Case Studies}
\label{sec:case_study}

\autoref{fig:cases} shows glyph-based visualizations reproduced with \toolname{}.
All of them are reproductions of award-winning visualizations from the website \textit{Information is Beautiful}.
We use two examples to showcase the creation process.



\begin{table}[ht]
\centering
\scriptsize
\setlength{\tabcolsep}{2pt}
\caption{Inputs and GDSL operations for constructing the snowflake glyph. \textbf{[Figma]} denotes canvas manipulation, \textbf{[NL]} a natural-language instruction, and \textbf{[Prototype]} an inferred operation.}

\begin{tabular}{p{0.34\columnwidth}|p{0.58\columnwidth}}
\hline
\textbf{User Input / Action} & \textbf{Parsed GDSL Operation} \\
\hline
\textbf{[Figma]} Draw 2-3 polylines in canvas & 
\texttt{CreateBasic("polyline1", "polyline", \{points, stroke, width\}, \{origin\})} \\
\hline
\textbf{[Prototype]} Infer a repeated structure &
\texttt{CreateRepeater(}\\
& \quad\texttt{"MultiPolyline", "polyline1",}\\
& \quad\texttt{"cartesian", 7, "uniform")} \\
\hline
\textbf{[Figma]} Draw a vertical line & 
\texttt{CreateBasic("line1", "line", \{start, end, stroke\}, \{origin\})} \\
\hline
\textbf{[Figma]} Combine vertical line and repeated polylines into a branch & 
\texttt{CreateCompositor(}\\
& \quad\texttt{"branch1", ["line1",}\\
& \quad\texttt{"MultiPolyline"], ...)} \\
\hline
\textbf{[NL]} \textit{``rotate and copy the branch 6 times''} & 
\texttt{CreateRepeater(}\\
& \quad\texttt{"snowflake", "branch1",}\\
& \quad\texttt{"polar", 6, "uniform")} \\
\hline
\textbf{[Figma]} Draw a hexagon & 
\texttt{CreateBasic("hexagon1", "polygon", \{sides: 6, stroke, width\}, \{origin\})} \\
\hline
\textbf{[Figma]} Combine the hexagon and the 6 branches together & 
\texttt{CreateCompositor(}\\
& \quad\texttt{"finalGlyph", ["snowflake",}\\
& \quad\texttt{"hexagon1"], ...)} \\
\hline
\end{tabular}
\label{tab:snowflake_operations}
\end{table}

\begin{figure}[h]
    \centering
    \includegraphics[width=.9\columnwidth]{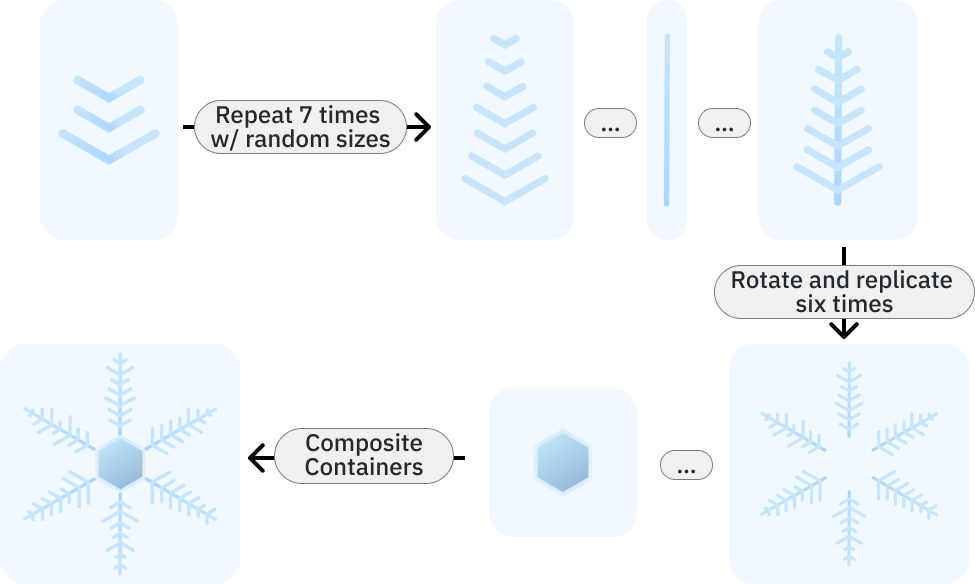}
    \caption{Creating a snowflake glyph using \toolname{} includes four steps: creating a repeater container, compositing with the vertical line, creating a repeater container for branches, and compositing with the hexagon.}
    \label{fig:casesnow}
\end{figure}

\textbf{Case 1: snowflake visualization.}
The visualization in \autoref{fig:cases} (d) originates from Twitter discussion intensity data regarding different actors across six seasons of the television series \textit{Game of Thrones}~\cite{TwitterInteractive2025}. 
Each snowflake represents an actor.
The six branches of a snowflake correspond to an actor's intensity over the six seasons.
Each branch consists of several polylines, further detailing the mention frequency for each episode.
The snowflake is authored in two stages. During design exploration, users first generate index-dependent placeholder values to inspect the repeated structure without entering the final dataset. These placeholder values can later be replaced by the character's actual Twitter mention frequencies, as in \autoref{fig:cases}(d). The
polylines within each branch correspond to the episodes of that season in
chronological order. Suppose the first user-drawn polyline has length $L_1$ and
corresponds to the first nonzero reference value $x_1$, \toolname{} treats this
polyline as the base visual unit with a scale factor of 1.0. A subsequent
value $x_i$ is assigned the relative scale factor $x_i/x_1$, producing
$L_i=L_1(x_i/x_1)$. Thus, the absolute lengths depend on the initial geometry
drawn by the user, whereas their relative lengths encode the mention
frequencies.
The central hexagon uses color as a categorical identifier for the character;
its size does not encode a data value. The overall glyph size is not
independently mapped to another variable, but results from the user-defined
base geometry and the relative scaling of the constituent polylines.
Traditionally, designing such snowflake glyphs requires manually editing each polyline, which is labor-intensive.
The construction complexity increases significantly when polylines vary in size and users need to draw or edit dozens of polylines.
As illustrated in \autoref{fig:casesnow}, with \toolname{} users can construct a complete snowflake visualization through a few steps (in \autoref{tab:snowflake_operations}):
\textit{Step 1. Initial Pattern Creation.} Users only need to draw 2-3 polylines in the Figma canvas, after which \toolname{} infers the user's intention to create an \repeaterContainer{RC} with \basicContainer{BC} polylines through Cartesian repetition and scaling.
Users can modify the number of repetitions, set to seven in this example, and inspect the rendered structure in the preview panel.
\textit{Step 2. Create a branch.}
When users add a vertical line, \toolname{} creates a basic container for the line and composites the vertical line and the repeated polylines into a \compositorContainer{CC}, a snowflake branch.
\textit{Step 3. Radial repetition.} Through the dialogue panel, users can specify \textit{``rotate and copy the branch 6 times''}. 
\toolname{} takes the NL instruction and current GDSL instance as input and outputs the operation for creating an \repeaterContainer{RC} in a polar coordinate system with 60-degree intervals repeated 6 times.
The results are also shown in the dialogue panel, with textual explanation and ambiguity widgets.
Users can adjust parameters in the preview panel and the dialogue panel to optimize placement, for example, adjusting the rotation center.
\textit{Step 4. Add a hexagon.} Users add a hexagon to the snowflake structure, then replicate this snowflake and encode data by assigning different colors to the hexagons in each instance.

\begin{figure}[h]
    \centering
    \includegraphics[width=0.8\columnwidth]{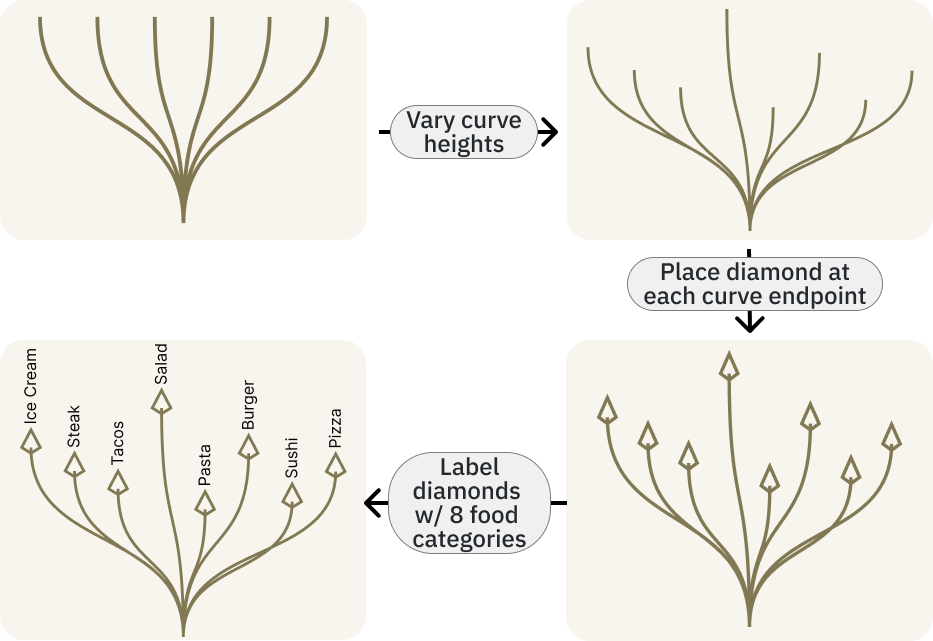}
    \caption{The process of designing protein visualization.}
    \label{fig:caseprotein}
\end{figure}

\textbf{Case 2: protein visualization.}
The design~\cite{McCandless2023Protein} in \autoref{fig:cases} (b) is an award-winning design from the \textit{Information is Beautiful Award}.
The visualization employs a plant-inspired structure where curved lines at the base connect to represent proteins of foods, with varying heights indicating different protein concentrations.
Foods in the same category are clustered together as a ``plant''.
As shown in \autoref{fig:caseprotein}, using \toolname{}, users can create this visualization by creating a cluster of lines containing 2-3 curves.
After analysis, \toolname{} determines that these curves were generated using different $x$-direction scaling factors following an arithmetic sequence.
\toolname{} reproduces the graphic by creating an \repeaterContainer{RC} with different $x$-scales.
\toolname{} enables users to specify different heights to the curves by using an NL instruction \textit{``give curves with different heights, randomly varying between 0.8 and 1.5''}.
This step applies random $y$-direction scaling to the curves, effectively encoding protein concentration values.
Subsequently, users can draw a diamond on the Figma canvas and enter \textit{``place this diamond above each curve''} in the dialogue panel.
\toolname{} uses a \funcname{CreateCompositor} operation to incorporate this diamond element (a \basicContainer{BC}) into the GDSL instance and positions it above each curve, establishing a spatial relationship.
Users can then request vertical text placement above each diamond through an NL instruction.
The text labels represent various food names, produced by a data encoding operation with a list of food names responded by the MLLM.
This way, the corresponding textual information is encoded into the visualization.
\autoref{fig:caseprotein} demonstrates this process through four stages: initial curve generation, height variation application, diamond placement above each curve, and finally, the addition of textual labels positioned vertically above each diamond.
After completing these steps, users can further enhance the visualization by creating filled diamonds with different sizes and colors to create the complete structure shown in \autoref{fig:cases} (b).

\subsection{Translation Benchmark}
\label{sec:quantitative_evaluation}

We evaluated how accurately four MLLM-based translators convert NL instructions into supported GDSL operations. We first extracted representative operation scenarios from the glyph-authoring cases and then designed 25 instructions for each of four visual-authoring operations: repeating content, combining an existing GDSL component with a target, modifying a fixed parameter, and encoding data. The resulting benchmark contains 100 unique instructions, each paired with an input GDSL instance and one or more accepted operation outputs. It tests translation within this balanced operation space rather than representing the full distribution of naturally occurring design requests. \revision{The supplementary material contains the parsing prompt (Appendix~A) and the benchmark and release-file index (Appendix~B).}

%

\begin{table}[t]
\centering
\caption{NL-to-GDSL translation results on the 100-case benchmark. Latency is client-side wall-clock time from one recorded response per case and is reported descriptively.}
\label{tab:llm_performance}
\resizebox{\linewidth}{!}{

\begin{tabular}{lcccc}
\hline
\shortstack[c]{\textbf{Model}\\ \textbf{ }}
& \shortstack{\textbf{Category Acc} \\ \textbf{(\%)}} 
& \shortstack{\textbf{Param Match} \\ \textbf{(\%)}} 
& \shortstack{\textbf{Perfect Match} \\ \textbf{(\%)}} 
& \shortstack{\textbf{Latency} \\ \textbf{(s)}} 
\\ \hline
GPT-5.4 & 100.00 & 92.00 & 82.00 & 3.90 \\
GPT-5.1 & 100.00 & 95.67 & 89.00 & 3.74 \\
GPT-4o & 100.00 & 92.33 & 84.00 & 2.27 \\
GPT-4o-Mini & 99.00 & 90.33 & 79.00 & 2.41 \\ \hline
\end{tabular}
}
\end{table}

\textbf{Benchmark Construction.}
Three annotators specified the expected operation type, target element, and operation-specific fields for each instruction. They resolved disagreements through discussion. When multiple GDSL outputs were judged semantically equivalent, the case-specific ground truth records the accepted alternatives rather than requiring textual identity across the complete serialization. We tested the snapshots \texttt{gpt-5.4-2026-03-05}, \texttt{gpt-5.1-2025-11-13}, \texttt{gpt-4o-2024-08-06}, and \texttt{gpt-4o-mini-2024-07-18} using the same prompt, a temperature of 0, and a maximum output length of 500 tokens. Each model produced one recorded response for each case.

\textbf{Evaluation Metrics.}
Category Accuracy requires an exact match among the four operation types. Parameter Matching is the macro-average of the scored operation-specific fields for each case. Perfect Match requires the operation type and every scored field to match an accepted ground truth. For example, if a \funcname{ModifyParams} prediction selects the correct operation, attribute, and value but the wrong target ID, it is correct for Category Accuracy, matches two of the three scored parameters, and is incorrect for Perfect Match. Missing or unparseable outputs count as incorrect. Mean Recorded Latency is client-side wall-clock time; because network and provider load vary and each case was run once, it is descriptive rather than a controlled performance comparison.

\textbf{Results and Analysis.}
\autoref{tab:llm_performance} shows that the models almost always selected the intended operation category (99--100\%). Parameter matching ranged from 90.33\% to 95.67\%, while perfect match ranged from 79\% to 89\%. GPT-5.1 obtained the highest parameter matching and perfect-match results; GPT-4o had the lowest recorded mean latency. The remaining mismatches primarily involved inferred repetition counts or rules, selected target IDs and composition relations, and data-expression fields. These results support the feasibility of constrained NL-to-GDSL translation for the tested instructions, but they do not establish performance for unrestricted design language or statistically compare model latency.

\subsection{User Study}

The user study examines the usability of \toolname{}, how participants perceive and use its integrated workflow combining natural-language input and direct manipulation, and their perceived efficiency when authoring glyph-based visualizations.
Specifically, the study addresses the following two research questions (RQs):
\textit{\textbf{RQ1}: Can users easily learn and use \toolname{} to author glyph-based visualizations without experiencing a steep learning curve?
\textbf{RQ2}: How do participants perceive and use the integrated workflow combining natural-language input and direct manipulation when expressing and refining glyph-design intentions?}
We additionally report participants' perceived efficiency as a descriptive study theme, rather than treating it as a research question about measured time or effort reduction.

\begin{figure*}[ht]
    \centering
    \includegraphics[width=.9\textwidth]{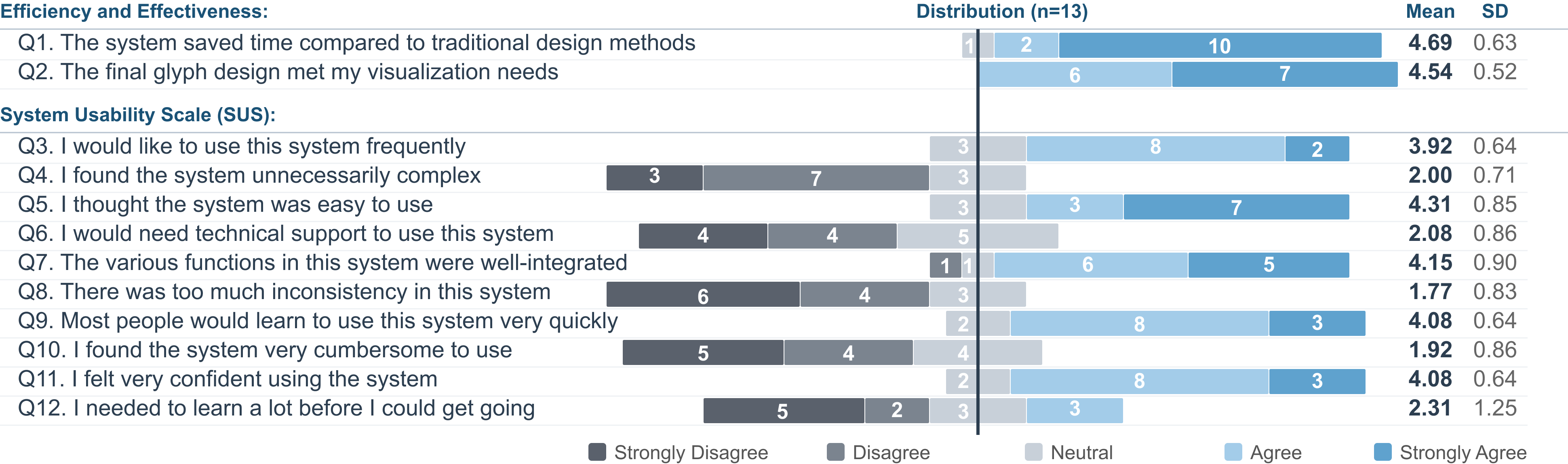}
    \caption{Questionnaire responses from the user study. Q1--Q2 are study-specific items, and Q3--Q12 comprise the System Usability Scale. Bars show the distribution of 5-point Likert responses; labels report the mean and standard deviation.}
    \label{fig:userstudy}
\end{figure*}

We conducted a user study with 13 participants (P1 to P13; 4 males, 9 females) from the fields of visualization and infographic design.
Specifically, six participants were from a school of design, majoring in infographic design.
The remaining were Master's or Ph.D. students specializing in data visualization.
Ten participants reported being relatively familiar with glyph design, with six of them (P2, P4, P9, P10, P11, P12) reporting high familiarity with glyph design.
All participants were proficient in designing infographics using tools like \textit{Figma}, and each had at least one year of experience in infographic design or visualization development.
\revision{During the user study, \toolname{} used GPT-4o (\texttt{gpt-4o-2024-08-06}) to translate participants' natural-language instructions into GDSL operations.}
Our user study followed the structured process outlined below:
\begin{itemize}[noitemsep,topsep=2pt,parsep=0pt,partopsep=0pt]
\item \textbf{Prototype Introduction.}
Participants were first provided with an overview of \toolname{}, including its integration with Figma.
We demonstrated \toolname{}'s features, with emphasis on the two supported interaction modes: NL instructions and direct manipulation on the Figma canvas.

\item \textbf{Task Execution.}
Participants completed two activities: reproducing one glyph-based visualization from \autoref{sec:case_study} to experience the supported authoring operations, and using the available features to create a visualization of their own interest.
Each participant was allocated approximately one hour to complete all study activities.

\item \textbf{Questionnaire.}
Immediately after completing the design tasks, participants filled out a questionnaire using a 5-point Likert scale (1: Strongly Disagree, 5: Strongly Agree). As shown in \autoref{fig:userstudy}, the reported questionnaire items comprise two parts: (1) two study-specific items on efficiency and effectiveness and (2) the standard 10-item \textit{System Usability Scale (SUS)}, evaluating usability, learnability, and perceived complexity.


\item \textbf{Interview.}
We conducted interviews with each participant to gain deeper insights into their user experience, challenges encountered, and suggestions for improvement.

\end{itemize}


The questionnaire results are shown in \autoref{fig:userstudy}.
The questionnaire results are reported alongside corresponding observations from the post-study interviews.

\textbf{Prototype Usability and Learnability.}
We calculated the SUS score from Q3--Q12; the mean score was 76.15, indicating good usability. Responses indicated that participants generally found the prototype easy to use (Q5: $\mu=4.31$, $\sigma=0.85$) and disagreed that it was unnecessarily complex (Q4: $\mu=2.00$, $\sigma=0.71$) or cumbersome (Q10: $\mu=1.92$, $\sigma=0.86$).
Furthermore, participants agreed that they would learn to use it quickly (Q9: $\mu=4.08$, $\sigma=0.64$) and felt confident during its use (Q11: $\mu=4.08$, $\sigma=0.64$).
In the interviews, P1, P4, and P8 associated their ease of adoption with the prototype's integration into their Figma workflow.

\textbf{Experience of the Integrated Multimodal Workflow.}
The SUS integration and consistency items indicated that participants generally perceived the prototype's functions as well integrated (Q7: $\mu=4.15$, $\sigma=0.90$) and reported a low perceived level of inconsistency (Q8: $\mu=1.77$, $\sigma=0.83$). These items assess the consistency of the prototype as a whole; they do not compare the two interaction modalities.
The study-specific item on fulfillment of visualization needs received a positive rating (Q2: $\mu=4.54$, $\sigma=0.52$).
P3 described the dialogue panel as a natural complement to direct canvas manipulation. P1 reported that NL instructions were particularly useful for repetitive tasks because one instruction could replace multiple manual operations.
P3 reported using ambiguity widgets to adjust inferred parameters, and P6 used the preview to inspect those adjustments. P9 and P11 referred to corrective feedback when an instruction could not be parsed.
The study examined participants' experience of the integrated workflow and did not isolate the respective contributions of natural-language input and direct manipulation. A controlled comparison of the two modalities and their combination is left for future work.

\textbf{Efficiency.}
The time-saving item received the highest mean among the questionnaire items (Q1: $\mu=4.69$, $\sigma=0.63$).
In the interviews, most participants estimated that they completed the assigned glyph-authoring tasks with \toolname{} in approximately 10--15 minutes. \revision{P12 estimated from personal experience that producing a polished visualization through a traditional collaborative workflow involving design and development could take approximately two weeks.} Because the latter estimate covers a broader process that may include ideation, refinement, and implementation, these self-reported durations are not directly comparable. We therefore interpret them as qualitative evidence of participants' perceived efficiency rather than as a measured speedup; no controlled baseline study was conducted.

\textbf{Suggestions for Improvement.}
Participants also identified limitations and suggested improvements.
One suggestion was to include enhanced data processing capabilities, such as data cleaning and transformation functions, which could reduce dependence on external tools before importing data (P5).
The current prototype leaves data cleaning and transformation outside the glyph specification, so this suggestion would require an additional upstream data-processing component.
Participants also suggested providing more educational resources, including example cases, reusable semi-finished templates, and a community-driven template repository to support learning and onboarding (P3, P9, P11).
A participant (P12) expected higher-level interaction mechanisms, allowing them to describe design intentions at a conceptual level rather than constructing visualizations step by step through low-level operations.
Several participants suggested generating initial visual structures from natural-language descriptions instead of first drawing them in Figma. P13 also requested clearer progress indicators for long-running operations.

\Needspace{10\baselineskip}
\section{Discussion}


\textbf{\revision{Mappings and reuse.}}
\toolname{} helps bridge user intent and formal specifications of glyph-based visualizations through both natural language input and direct manipulation. Our case studies show that GDSL can represent a range of structures with repeated and hierarchical organizations, illustrating the flexibility of its composition model.
Notably, the prototype supports nested data structures beyond simple tabular data, enabling more complex and diverse visualization scenarios.
\revision{GDSL supports both one-to-one and many-to-one mappings. A basic container can map different variables to different attributes of one object, such as size and color (one-to-one). A repeater container can map multiple variables from one data record to the same attribute type on different repeated elements, such as the lengths of different petals in one glyph (many-to-one).}
Although \toolname{} is primarily demonstrated with standalone and infographic-style examples, it can also support interactive visual analytics workflows.
Its D3.js-based rendering pipeline exports authored glyphs as reusable code that can be rebound to new data in application-specific interfaces, including small multiples, overview displays, dashboards, and map overlays.

\textbf{\revision{System limits.}}
However, the current GDSL does not natively support automatic layout computation, and some cases therefore require additional processing outside the specification itself.
For example, for node link diagrams based on force directed layouts, \toolname{} allows users to bind data directly to node and link coordinates, but it does not support computing these positions within the specification. We treat such computation as an extra independent functionality that is decoupled from the visual design specification, since it depends on upstream data processing rather than on visual structure itself.
This limitation reflects the broad variety of layout methods, which makes comprehensive built-in support difficult in the current prototype.
As a result, the present implementation focuses on direct data mapping rather than automatic layout generation.
\revision{When multiple data variables jointly determine a single visual attribute, users can also precompute a derived value outside GDSL and then bind that value to the attribute.}
Importantly, we note that data cleaning and wrangling are not the focus of glyph design itself. Such preprocessing tasks involve decisions about data semantics and transformation pipelines, which fall outside the scope of glyph specification.

\textbf{\revision{Design flexibility.}}
\revision{\toolname{} helps users turn design ideas into glyphs they can inspect and refine without writing code from scratch. Because glyphs often use metaphors and context-specific encodings, users can define data-to-visual bindings that suit their designs. The prototype leaves these choices to users without enforcing strict validation, which can lead to unsuitable encodings, especially for users with limited experience. We plan to add warnings that explain potential problems and suggest alternatives while letting users proceed. For example, a warning could explain that assigning nominal data to a continuous gradient implies an order and suggest a categorical palette.}

\textbf{\revision{Study limitations.}}
\revision{All 13 participants had taken part in an evaluation of \toolname{} about one year earlier. In this study, they completed new glyph-authoring tasks and immediately filled out the questionnaire, including the ten-item SUS. Prior experience with \toolname{} may have influenced their learnability ratings (RQ1) and SUS scores.}

\section{Conclusion and Future Work}

We presented \toolname{}, an interactive prototype for authoring glyph-based visualizations. Its GDSL represents their structures as hierarchies of basic, repeater, and compositor containers, and its operation management mechanism restricts MLLM translations to five inspectable GDSL operations.
The case studies show how this representation reconstructs the repeated and hierarchical structures of the examined glyph-based visualizations. The translation benchmark evaluates whether different models identify the intended operations and parameters, and the user study reports a mean SUS score of 76.15. The reported time estimates reflect perceived efficiency rather than a controlled comparison. These findings are limited to the examined corpus, benchmark instructions, participants, and tasks.
The prototype does not compute derived layouts, perform data cleaning, or validate the semantic appropriateness of data mappings. Future work will investigate them and evaluate \toolname{} against instrumented baseline workflows.


\section*{Acknowledgment}
This project is supported by the Ministry of Education, Singapore, under its Academic Research Fund Tier 1 (NTU Tier 1: RG104/25) and Academic Research Fund Tier 2 (Proposal ID: T2EP20222-0049). Any opinions, findings and conclusions, or recommendations expressed in this material are those of the author(s) and do not reflect the views of the Ministry of Education, Singapore.

\bibliographystyle{abbrv-doi-hyperref}

\bibliography{main}

\begin{IEEEbiography}[{\includegraphics[width=1in,height=1.25in,clip,keepaspectratio]{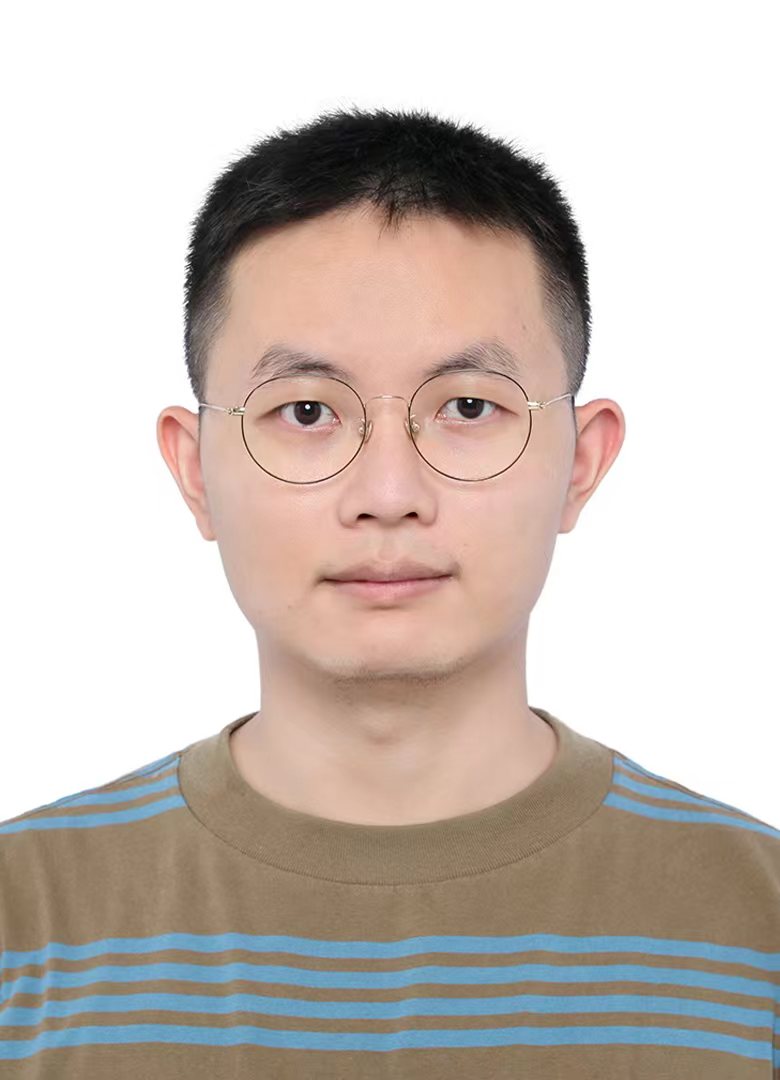}}]{Can Liu} is an assistant professor at the Central University of Finance and Economics.
He was previously a research fellow in the College of Computing and Data Science at Nanyang Technological University.
He received his Ph.D. in Intelligent Science (2023), B.Sc. in Computer Science (2018), and B.Ec in Economics (2018) degrees from Peking University.
His research interests lie in the field of deep learning-driven visualization, especially intelligent interaction for visualization. Refer to \url{https://vvoliucano.github.io} for more details.
\end{IEEEbiography}

\begin{IEEEbiography}[{\includegraphics[width=1in,height=1.25in,clip,keepaspectratio]{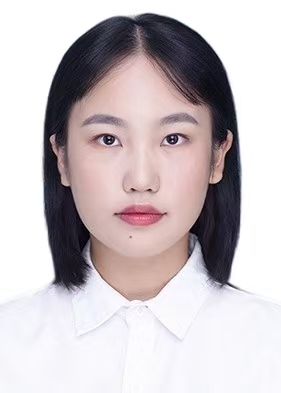}}]{Shiwei Chen}
is a master student at the  College of Computing and Data Science, Nanyang Technological University, Singapore. Her research interests include human--AI interaction and scientific visualization. She is currently exploring AI-assisted approaches to support creativity in visual representation.
\end{IEEEbiography}

\begin{IEEEbiography}[{\includegraphics[width=1in,height=1.25in,clip,keepaspectratio]{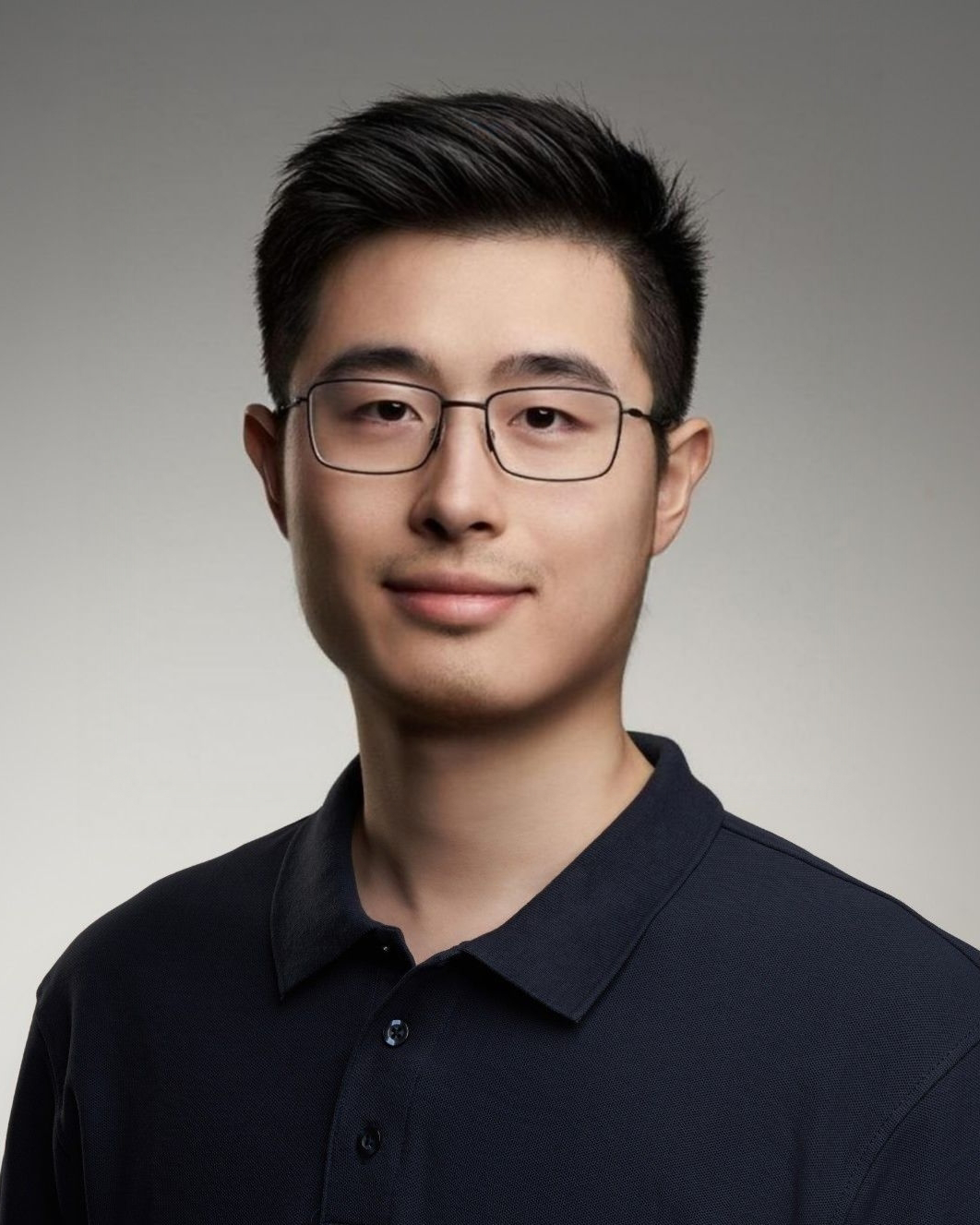}}]{Zhibang Jiang}
received a B.A. degree in journalism from Communication University of Zhejiang and an M.S. degree in data visualization from Parsons School of Design. His design interest is human-centered design for data-intensive services. His design footprints include tenures at leading organizations such as Providence, Tencent, Alibaba, Peking University, and GroupM, and collaborations with partners such as Nike, and the Smithsonian. He has shaped the data experience for millions of users.
\end{IEEEbiography}

\begin{IEEEbiography}[{\includegraphics[width=1in,height=1.25in,clip,keepaspectratio]{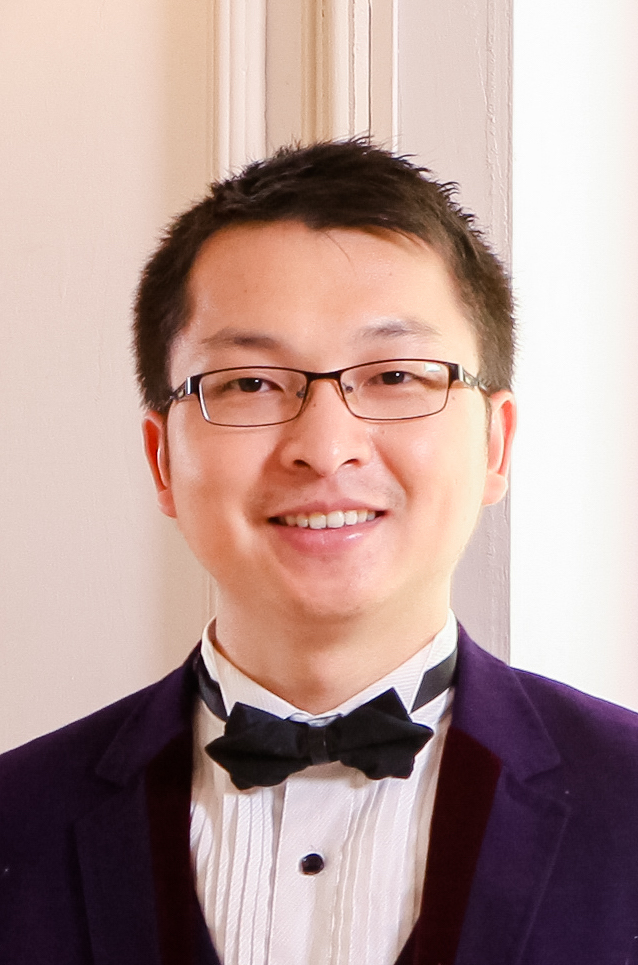}}]{Yong Wang}
is currently an assistant professor in the College of Computing and Data Science, Nanyang Technological University. Before that, he worked
as an assistant professor at Singapore Management University from 2020 to 2024. His research interests include data visualization, HCI, and human-AI collaboration, with an emphasis on their application to FinTech, quantum computing and online learning.
He obtained his Ph.D. in Computer Science from Hong Kong University of Science and Technology. He received his B.E. and M.E. from Harbin Institute of Technology and Huazhong University of Science and Technology, respectively. For more details, please refer to \url{http://yong-wang.org}.

\end{IEEEbiography}

\clearpage



\end{CJK}     
\end{document}